\documentclass[
 reprint,
superscriptaddress,
nofootinbib,
bibnotes,
 amsmath,amssymb,
 aps,
prb,
]{revtex4-2}
\usepackage[normalem]{ulem}
\usepackage[plainpages = false, pdfpagelabels, 
                 bookmarks,
                 bookmarksopen = true,
                 bookmarksnumbered = true,
                 breaklinks = true,
                 linktocpage,
                 colorlinks = true,
                 linkcolor = blue,
                 urlcolor  = blue,
                 citecolor = blue,
                 anchorcolor = green,
                 hyperindex = true,
                 hyperfigures
                 ]{hyperref} 
\usepackage{verbatim}
\usepackage{enumerate}
\usepackage{makecell, multirow}
\usepackage{tikz}
\usetikzlibrary{patterns}
\usepackage{slashed}
\usepackage{braket}
\usepackage{bbold}
\usepackage{bm}
\usepackage{booktabs}
\usepackage[caption=false,singlelinecheck=off]{subfig}
\usepackage{blindtext}
\usepackage{bbm}
\usepackage{graphicx}
\usepackage{dcolumn}
\usepackage{bm}
\usepackage[mathlines]{lineno}
 \usepackage{esint}
\definecolor{ballblue}{rgb}{0.13, 0.67, 0.8}

\def \p {\partial}
\def \be {\begin{equation}}
\def \ee {\end{equation}}

\begin{document}

\title{Viscous flow through a finite-width slit: Boundary conditions and dissipation}

\author{Daniil Asafov}
 \affiliation{National Research University Higher School of Economics, 101000 Moscow, Russia}

 \author{Valentin Kachorovskii}
 \email{kachor.valentin@gmail.com}
  \affiliation{Ioffe Institute, 194021 
 St.~Petersburg, Russia}

 \author{Konstantin Tikhonov}
  \email{tikhonov@itp.ac.ru}
 \affiliation{L. D. Landau Institute for Theoretical Physics, 142432 Chernogolovka, Russia}

\author{Gu Zhang}
 \email{zhanggu@baqis.ac.cn}
 \affiliation{Beijing Academy of Quantum Information Sciences, Beijing 100193, China}

\begin{abstract}
We study the hydrodynamic viscous electronic transport in a two-dimensional sample  separated    into  two  semi-infinite planes by a  one-dimensional infinite  barrier. The semi-infinite planes are electrically connected via the finite-size slit in the barrier.
We calculate the current through the slit assuming finite voltage drop between the planes and neglecting  disorder-induced Ohmic resistance, so dissipation and resistance are purely viscosity induced.
We find that the only solution to the Stokes equation in this geometry, which yields a finite dissipation at finite resistance (and, hence, is not self-contradictory), is the one that fulfills both the no-stress and no-slip boundary conditions simultaneously. As a remarkable consequence,  the obtained velocity profile satisfies the so-called  ``partial-slip'' (Maxwell)  boundary condition for any value of the slip length, which drops out from all final results. We also calculate the electronic temperature profile for the small and large heat conductivity, and find asymmetric (with respect to the barrier) temperature patterns in the former case.

\end{abstract}

\date{\today}
\maketitle

\section{Introduction}

Rapid developments of nanotechnology in recent years led to the fabrication of ultra-clean ballistic systems, where  disorder scattering is negligible.     As a result, we are witnessing a boom of interest 
in electronic transport in the viscous hydrodynamic regime, where electron motion is governed by the Stokes equation. In the last decade, various exotic and unique features of electronic hydrodynamics were predicted theoretically (for review, see Refs.~\cite{NarozhnyAnnalenderPhysik17,LucasFongJPhys18,NarozhnyAofPhys19,PoliniGeimPhysicsToday20}). These include super-ballistic charge transport (with conductance higher than the ballistic conductance) through a slit~\cite{GuoPNAS17}, as well as an abrupt viscosity-induced jump in the electric potential at the metallic contacts~\cite{ShavitShytovFalkovichPRL19}. These predictions, which greatly contrast with common knowledge on transport in disordered systems, have inspired a great amount of  recent experimental investigations~\cite{BandurinScience16, CrossnoScience16,MollScience16,GhahariPRL16,KumarNatPhys17,BandurinNatureComm18,BraemPRB18,AlexandreNPJQM18,LevinBakarovPRB18,GoothNatCom18,BerdyuginScience19,GallagherScience19,SulpizioNatPhys19,EllaNatNano19,MarkNature20,RaichevPRB20,GusevSciRep20,VoolX2020,GenursX2020,KimNatCom20,JenkinsX20,GuptaPRL21,GusevPRB21,ZhangShurMichaelJounalAppPhys21,JaouiNatCom21,KrebsX21,KumarX21,SamaddarNanoLett21,AmitZeldovNature22}.
To date, most of these studies focus on the electric properties including the electric potential and the velocity profiles.
Most recently, the fast development of the SQUID-on-tip (tSOT) \cite{FinklerNanoLett10,Vasyukov,HalbertalNat16,HalbertalScience17,Zeldov2019,UriNatPhys20,AnahoryNanoscale20,AmitNature21,AmitZeldovNature22}
and the cryogenic quantum magnetometry \cite{VoolX2020} techniques have enabled another intriguing direction of research---the high-resolution detection of the temperature distribution in the hydrodynamic regime.

On the theoretical side, solutions in the hydrodynamic regime rely heavily on the choice of boundary conditions.
The physics behind these conditions is related to the friction force at the boundary (for a  detailed discussion see Ref.~\cite{EgorJorgPRB19} and references therein). Very strong friction corresponds to the boundary with diffusive scattering and  is usually described by the  no-slip boundary condition with zero velocity along the boundary 
\begin{equation}
\text{no-slip:}\quad    v_\parallel |_\text{boundary} = 0.  
\label{eq:no-slip}
\end{equation}
For  a  smooth boundary where friction is absent, one usually uses the no-stress boundary condition, 
with zero shear stress at the boundary, 
\begin{equation}
\text{no stress:}\quad 
    \partial_{r_\perp} v_\parallel |_\text{boundary} = 0,
\label{eq:no-stress}
\end{equation}
conditions. 
Here $r_\perp$ and $v_\parallel$ refer to the direction perpendicular to the boundary and the velocity along the boundary, respectively. 
It was proposed~\cite{TorrePoliniPRB15,FalkovichLevitovPRL17} that these two boundary conditions can in general be unified by the ``partial-slip'' (Maxwell \cite{Maxwell}) boundary condition:

\begin{equation}
v_\parallel\Big|_\text{boundary} = \lambda \partial_{r_\perp} 
v_\parallel 
\Big|_\text{boundary}.
 \label{eq:boundary_condition}
\end{equation}
Physically, $\lambda$ can be interpreted as the slip length for particles at the boundary.
Typically,  solutions of the Stokes equation with boundary conditions Eqs.~\eqref{eq:no-slip} and \eqref{eq:no-stress} result in  distinct velocity profiles with different characteristics. A prime  example is the Poiseuille flow   which shows finite and zero resistance for  no-slip and no-stress conditions, respectively.          This implies that, generically, the system-specific  boundary conditions can be  probed   experimentally.
Recently,  there have also been theoretical debates on the validity of these two boundary conditions in graphene ~\cite{TorrePoliniPRB15,GuoPNAS17,EgorJorgPRB19}. It is also worth noting that the no-slip \eqref{eq:no-slip} and the no-stress
 \eqref{eq:no-stress} boundary conditions are frequently associated with the limits $\lambda\to 0$ and $\lambda\to \infty$ in Eq.~\eqref{eq:boundary_condition}, respectively. However, as we are going to show in this paper, this natural distinction may fail, since both sides of Eq.~(\ref{eq:boundary_condition}) may be zero.

More precisely, in this paper,  we demonstrate that, surprisingly, the no-stress and no-slip boundary conditions  can  co-exist, i.e., can be satisfied simultaneously in some specific geometries. This, in turn, means  that  the Maxwell  boundary condition, Eq.~\eqref{eq:boundary_condition}, is satisfied for any $\lambda$, and, consequently, the velocity profile does not depend on $\lambda$.

\begin{figure}
  \includegraphics[width=\linewidth]{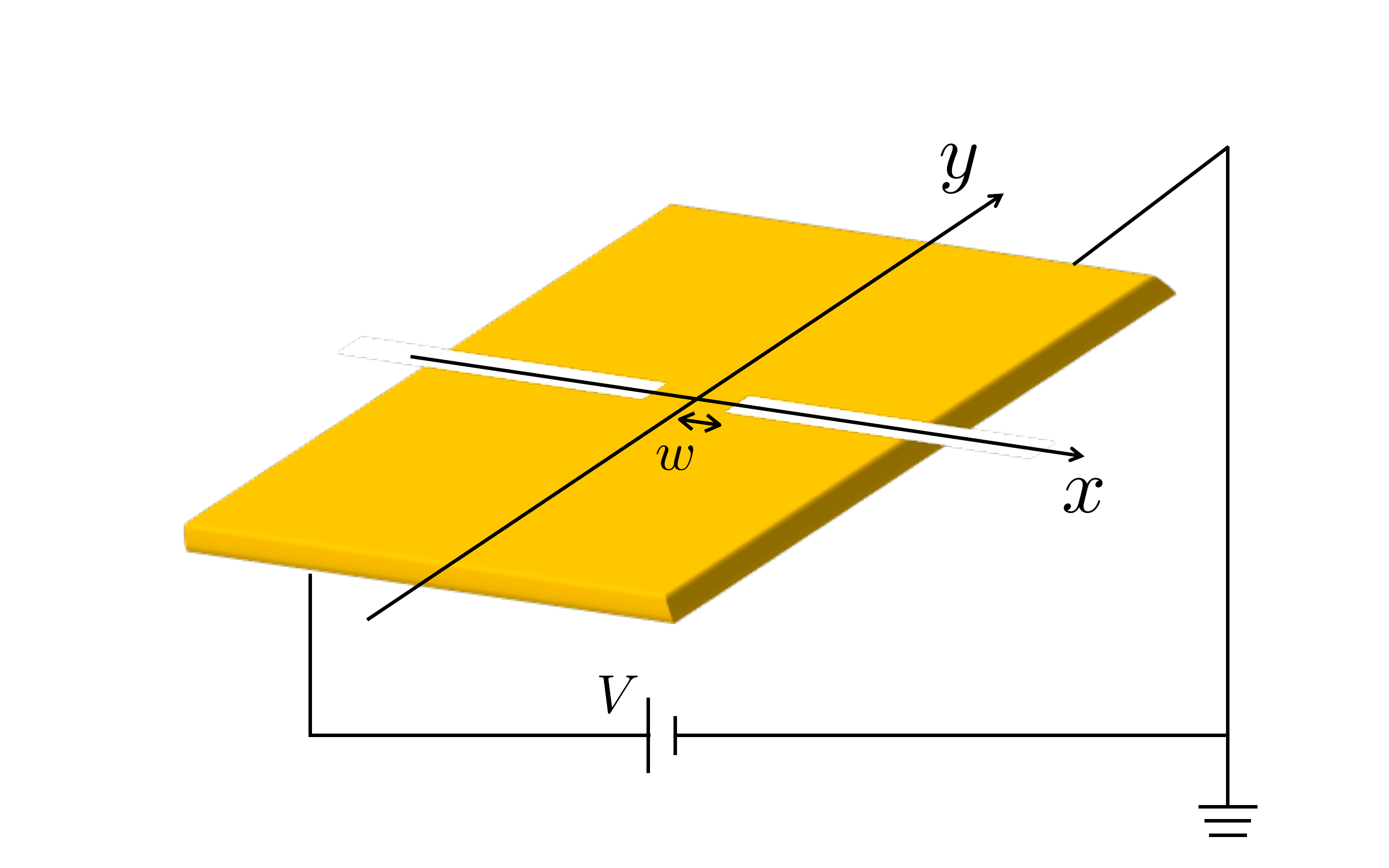}
  \caption{Schematics of the two-dimensional sample we consider. Charge transfer is only allowed through a slit located at $y = 0$ inside the segment from $x = -w$ to $x = w$.
  The rest of the sample at $y=0$ constitutes the impenetrable barrier (white region). The barrier thickness is infinitesimal in the $y$ direction. The sample is biased by a voltage $V$ between the $y < 0$ and the $y>0$ half-planes.}
  \label{fig:setup}
\end{figure}

We analyze   the velocity and temperature profiles in a two-dimensional (2D) sample where two semi-infinite planes are separated by the infinite barrier with a finite-size slit, as shown in Fig.\,\ref{fig:setup}.
Such a structure can be easily realized experimentally by using specific gate configurations.
Electronic flow through  similar systems and associated dissipation patterns have been studied  theoretically in   Ref.~\cite{RokniLevinson1995}  for disorder-dominated regime ignoring viscous effects. More recently, this system  has been addressed in the context of viscous electronic hydrodynamics  Refs.~\cite{FalkovichLevitovPRL17,GuoPNAS17,PershogubaYongGlazmanPRB20,LiKhodasLevchenkoPRB21,QiLucasPRB21}.

We neglect the disorder-induced sample resistance, and investigate the viscosity-controlled hydrodynamic velocity profile for fixed voltage drop across the system, i.e., between lines $y=-\infty$ and $y=\infty$. 
We show  that this condition  together  with     any standard boundary conditions, say ``no-slip'' or ``no-stress'', is not sufficient to uniquely find the velocity profile.   In particular, we  find  that  the ``no-stress''  boundary condition allows for a family of solutions with a finite    total current through the slit [see below solution,  Eq.~\eqref{general_noslip}].  We demonstrate  that the solution  can be uniquely determined only  after enforcing the total dissipation in the whole system to be finite. 
Remarkably, the  solution corresponding to finite dissipation 
satisfies simultaneously both the ``no-slip'' and ``no-stress'' boundary conditions, and, consequently,  satisfies
the partial-slip boundary condition, Eq.~\eqref{eq:boundary_condition}, 
for any value of $\lambda.$  
 This is our key result.

%%%%%%%%%%%%%%%%%%%%%%%%%%%%%%%%%%%%%%%%%%
%%%%%%%%%%%%%%%%%%%%%%%%%%%%%%%%%%%%%%%%%%

  The system under study has been discussed  previously. In Ref.\,\cite{PershogubaYongGlazmanPRB20}, a solution was found, which satisfies the no-stress condition \eqref{eq:no-stress}, but does not satisfy the no-slip condition \eqref{eq:no-slip}. As we demonstrate below, this solution yields a divergent total dissipation and, therefore, is nonphysical.  The  solution we obtained  is also not captured by the point-contact approximation used  in Ref.~\cite{FalkovichLevitovPRL17}.   Although  the  resistance in our solution   equals to  the ``superballistic'' resistance reported in Ref.\,\cite{GuoPNAS17}  for  the no-slip boundary conditions, and the velocity profile in the whole plane coincides  with Eq.~(S14)  of Ref.~\cite{GuoPNAS17}, the fact that this solution also satisfies the no-stress condition \eqref{eq:no-stress} was not mentioned there.

In the current paper, we also discuss the methodological aspects of the calculations. We particularly demonstrate a technical problem in taking the point-source limit (of Ref.~\cite{FalkovichLevitovPRL17}): For a system with 
an arbitrarily small slit size, one needs to take care about positions whose distances to one slit end point is much smaller than the slit size.

We  also solve the heat balance equation 
and find temperature  distribution.   
We demonstrate that the temperature is strongly  increased  near the slit end points. The shape of the
overheated regions   depends on heat conductivity. For  small heat conductivity, the heat transfer is dominated by convection and  these regions are strongly asymmetric with respect to the barrier separating the two half planes. In the limit of zero heat conductivity, the temperature diverges at the slit end points.   With increasing the heat conductivity, the   overheating decreases and the shape of the hot areas becomes more and more symmetric.

The rest of the  paper is organized as follows.
In Sec.~\ref{sec:model}, we describe the  model and hydrodynamic approach. In Sec.~\ref{sec:velocity}, we analyze the velocity profiles in light of the  finite-dissipation requirement. In Sec.~\ref{sec:finite_lambda}, we discuss the general approach for the case of  finite slip length. 
The electronic temperature profile is found in Sec.\,\ref{sec:temperature}.
Finally, we summarize our results in Sec.\,\ref{sec:summary}.
Technical details are presented in the Appendices.

\section{The Model}
\label{sec:model}

The model we consider consists of two clean homogeneous semi-infinite planes connected by a finite-width slit placed at the $x$ axis ($y = 0$ and $-w < x < w$), as shown in Fig.\,\ref{fig:setup}.
Particle transmission between these two planes is allowed only through this slit, but forbidden everywhere else.
We consider a viscous flow through the slit, which is generically described by the Stokes equation:
\begin{equation}
    \eta \nabla^2  \mathbf{v} = N e \mathbf{\nabla} \phi (\mathbf{r}).
    \label{eq:stokes_eq}
\end{equation}
Here, $\eta$ denotes the viscosity coefficients, $N$ is the particle density, $\mathbf{v} = (v_x, v_y)$ is the  fluid velocity vector, and $\phi$ is the electric potential, satisfying the Laplace equation $\Delta \phi = 0$.
We assume that the electron liquid is incompressible and quasi-neutral, so ${\rm div} \;\!\!\mathbf v=0$ and  $N=\text{const}.$  Both conditions are satisfied provided that  $s\gg v_{\rm F},~r_{\rm sc} \ll w$  where $s$ is the plasma wave velocity, $ v_{\rm F}$ is the Fermi velocity, and $r_{\rm sc}$ is the screening radius  (see discussion in Refs.~ \cite{Tikhonov2019,GuPRB21}).

Equation~\eqref{eq:stokes_eq} describes a sample with a negligible Ohmic resistance, where the driving field $-\nabla \phi$ is balanced by the viscous force.
From the energy conservation perspective, the driving power equals the dissipation due to viscosity, with the corresponding power density
\begin{equation}
    P(x,y) = \eta \left[{\left(\frac{\partial v_x}{\partial x} -\frac{\partial v_y}{\partial y}\right)^2 + \left(\frac{\partial v_y}{\partial x} + \frac{\partial v_x}{\partial y}\right)^2  }\right].
    \label{eq:dissipation_vis}
\end{equation}
We note that, once the velocity and/or the stress tensor at the wall are zero, there is no boundary contribution to the total energy change in the system \cite{ShavitShytovFalkovichPRL19}.

The non-equilibrium driving is included via the application of a voltage bias $V$ between the lower ($y<0$) and the upper ($y > 0$) half planes.
At positions far away from the slit, $r\equiv \sqrt{x^2 + y^2} \gg w$, the electric potential is given by
\begin{equation}
    \phi(\mathbf{r})|_{r \to\infty, y>0} = 0,\quad \phi(\mathbf{r})|_{r \to\infty, y<0} = V
    \label{eq:epotential_bc}
\end{equation}
in the two half-planes. The symmetry of the  problem implies the following relations for velocity components:
\be
\begin{aligned}
&v_x(x,y)= -v_x(x,-y)=-v_x(-x,y),
\\
& v_y(x,y)= v_y(x,-y)=v_y(-x,y).
\end{aligned}
\label{symmetry}
\ee

Equation~\eqref{eq:stokes_eq} has already been considered for the charge transport through a point contact \cite{FalkovichLevitovPRL17} or a finite-size slit \cite{GuoPNAS17,PershogubaYongGlazmanPRB20}.
After the introduction of the vorticity,
\begin{equation}
\mathbf{\omega} = [\mathbf{\nabla} \times \mathbf{v}]\cdot \hat{n},
\label{vorticity-def}
\end{equation}
the Stokes equation (\ref{eq:stokes_eq}) becomes
\begin{equation}
    [ \hat{n} \times \mathbf{\nabla} \omega (\mathbf{r}) ] = \frac{Ne}{\eta} \mathbf{\nabla} \phi(\mathbf{r}),
    \label{eq:viscous_stokes}
\end{equation}
where $\hat{n}$ is the unit vector perpendicular to the 2D plane. Taking the curl of Eq.~(\ref{eq:viscous_stokes}), one finds that the vorticity satisfies the Laplace equation: $\Delta\omega=0$.

Vorticity $\omega (\mathbf{r})$ and electric potential $\phi(\mathbf{r})$ satisfy the Cauchy-Riemann conditions in a complex plane with its real and imaginary axes directed along $x$ and $y$, correspondingly. This means that the combination of these two functions,
\begin{equation}
    f(z) = -\frac{Ne}{\eta} \phi (\mathbf{r}) + i \omega (\mathbf{r}),
    \label{fz-def}
\end{equation}
is a holomorphic function, with $z = x + iy$. In the no-stress case, the vorticity vanishes at the walls.

We thus have to find a complex function, holomorphic in the upper half plane, which has a prescribed imaginary part (vorticity) on some pieces of the boundary and the real part (electric potential) on other pieces of the boundary. Some particular solutions to this problem are delivered by the Keldysh-Sedov theorem\cite{TFKP}, see Appendix \ref{ks_app} for details. The function $f(z)$ can be found in the closed form
 \cite{PershogubaYongGlazmanPRB20}:
\begin{equation}
    f(z) = \frac{NeV}{2\eta} 
     \left[ -1 + \frac{\, z}{\sqrt{z^2-w^2}}\right],
    \label{eq:fz}
\end{equation}
where the cuts defining the square root $\sqrt{z^2-w^2}$ run along the screen. Note that with this convention,  $$\overline{\sqrt{z^2-w^2}} = -\sqrt{\bar {z}^2-w^2}$$ 
(throughout the paper, the bar denotes complex conjugation; e.g., 
$\bar{z}=x-iy$ is the complex conjugate of $z$).

The function $f(z)$ corresponds to the following velocity profile:
\begin{equation}
    u(z, \bar{z}) \equiv v_y + i v_x = \frac{eNV}{8 \eta i} \left[{\frac{z \bar{z} - \mathcal{C}}{\sqrt{z^2-w^2}} + \sqrt{\bar{z}^2-w^2}}\right],
    \label{eq:umain}
\end{equation}
with an undetermined constant $\mathcal C$ [expressions for $v_x$ and $v_y$ in the real form and in terms of the real coordinates $(x,y)$   can be found in Appendix~\ref{even_more_singular_app}].

One surprising feature of Eq.~\eqref{eq:umain}, derived in the no-stress limit, is that at first glance the velocity profile  is not uniquely determined. A closer inspection [see Appendix \ref{ks_app} for the analysis leading to Eq.~(\ref{eq:umain})] shows that this family exhausts all solutions with vorticity and electric potential
integrable in the vicinity of the slit endpoints. However, there exist other solutions to this problem, which have non-integrable vorticity, but finite total current passing through the slit, see Eq.~\eqref{general_noslip}, and infinitely many even more singular solutions, which support infinite current at a  finite voltage drop (see Appendix \ref{even_more_singular_app}). As we demonstrate below, the  integrability of velocity and vorticity  is not sufficient to  determine $\mathcal C.$  Importantly, to arrive at a unique $\mathcal{C}$, a finite total dissipation should also be enforced.

In Ref.~\cite{PershogubaYongGlazmanPRB20}, the coefficient $\mathcal{C}$ was fixed by comparing the velocity distribution given by Eq.~(\ref{eq:umain}) at $r\to\infty$ to that previously obtained for a point contact, $w \equiv 0$ \cite{FalkovichLevitovPRL17}, leading to
\begin{equation}
\label{ref_C}
\mathcal{C} = 2w^2.
\end{equation}
We will argue in what follows that this choice is problematic, as it leads to the solution that is internally inconsistent (in particular, not respecting the energy conservation law).
Moreover, we will show that the point-contact limit itself is not well-defined.

\section{Work, dissipation, and energy conservation}
\label{sec:velocity}

In this section, we take a closer look at the solution in Eq. (\ref{eq:umain}) obtained under the requirement of a vanishing stress tensor at the boundaries of the barrier and integrable vorticity. It is useful to express the dissipation power stemming from viscosity as a function of complex variables. Using Eq.~\eqref{eq:dissipation_vis}, we get
\begin{equation}
    P(z,\bar{z})= 4\eta\,\frac{\partial u}{\partial z} \cdot \frac{\partial  \bar{u}}{\partial \bar{z}}.
    \label{eq:Pzz}
\end{equation}
Another useful representation \cite{Hydrobook,Koh} for viscous dissipation is through vorticity, Eq.~(\ref{vorticity-def}):
\begin{equation}
    P(x, y) =\eta\mathbf{\omega}^2
    +2\eta~ {\rm div} \mathbf S,
\label{P1+P2}
\end{equation}
where
\be
\mathbf S= (\mathbf{v}\cdot\mathbf{\nabla})\mathbf{v}.
\label{S}
\ee
Therefore, the total dissipated power, $$\mathcal{P} = \iint P(x, y) dx dy,$$ can be separated into bulk and surface terms. The latter is determined by the flux of $\mathbf S$ through surface of the barrier.

In the vicinity of the slit endpoints, Eqs.~\eqref{eq:umain} and \eqref{eq:Pzz} give for the leading term of the dissipation power:
\begin{equation}
    P(z, \bar{z}) \propto \frac{(w^2-\mathcal{C})^2}{w} \left|\frac{\partial (z \pm w)^{-\frac{1}{2}}}{ \partial z}\right|^2 \propto \frac{(w^2- \mathcal{C})^2}{\rho ^3},
    \label{eq:near_slit_power}
\end{equation}
where $\rho = \sqrt{(x \pm w)^2 + y^2}$ denotes the distance from the slit endpoints.
For general choices of $\mathcal{C}$, Eq.~\eqref{eq:near_slit_power} diverges $\sim 1/\rho$ after the integration over $z$ and  $\bar{z}$, when $\rho$ approaches zero.
One can show that vorticity does not depend on $\mathcal C$ (see Appendix \ref{even_more_singular_app}).  Therefore, divergence in Eq.~\eqref{eq:near_slit_power} is fully determined by the surface term in Eq.~\eqref{P1+P2}.     One can also show (see Appendix \ref{even_more_singular_app}) that close to the slit end, the vector $\mathbf S$ scales as $\mathbf S \propto (\mathcal C - w^2)^2 ~\boldsymbol{ \rho}/\rho^3, $ so the right-hand side of Eq.~\eqref{eq:near_slit_power}  is proportional to ${\rm div} \mathbf S.$  Evidently, total dissipation power $\mathcal P$ also diverges.    As is clear from the above analysis, this  divergence     means  existence of unphysical energy flow through the slit ends.   

The divergence disappears if and only if $\mathcal{C} = w^2$.
This choice is different from that based on the matching of the velocity profile at large distances to the point-contact asymptotics, Eq. (\ref{ref_C}),
as was done starting with the no-stress requirement at the barrier in Ref.~\cite{PershogubaYongGlazmanPRB20}. Note that the velocity profile with $\mathcal{C} = w^2$ under the no-stress requirement is equivalent to the so called ``no-slip'' solution derived in Ref.~\cite{GuoPNAS17}, which assumed zero velocity at the barrier. It turns out that this no-slip solution is simultaneously characterized by zero stress at the boundary, and, thus, is a valid no-stress solution, too. 

For $\mathcal C=w^2,$ the surface term gives zero contribution to total dissipation, which can be shown by     integrating Eq.~\eqref{P1+P2}by parts. Hence, the total dissipated power reads as
\begin{equation}
\mathcal{P}
= \eta \iint \omega^2(x, y) dx dy, 
\qquad{\rm for} \quad \mathcal C=w^2 .
\label{eq:another dissipation expression}
\end{equation}

\begin{figure}
  \includegraphics[width=0.8\linewidth]{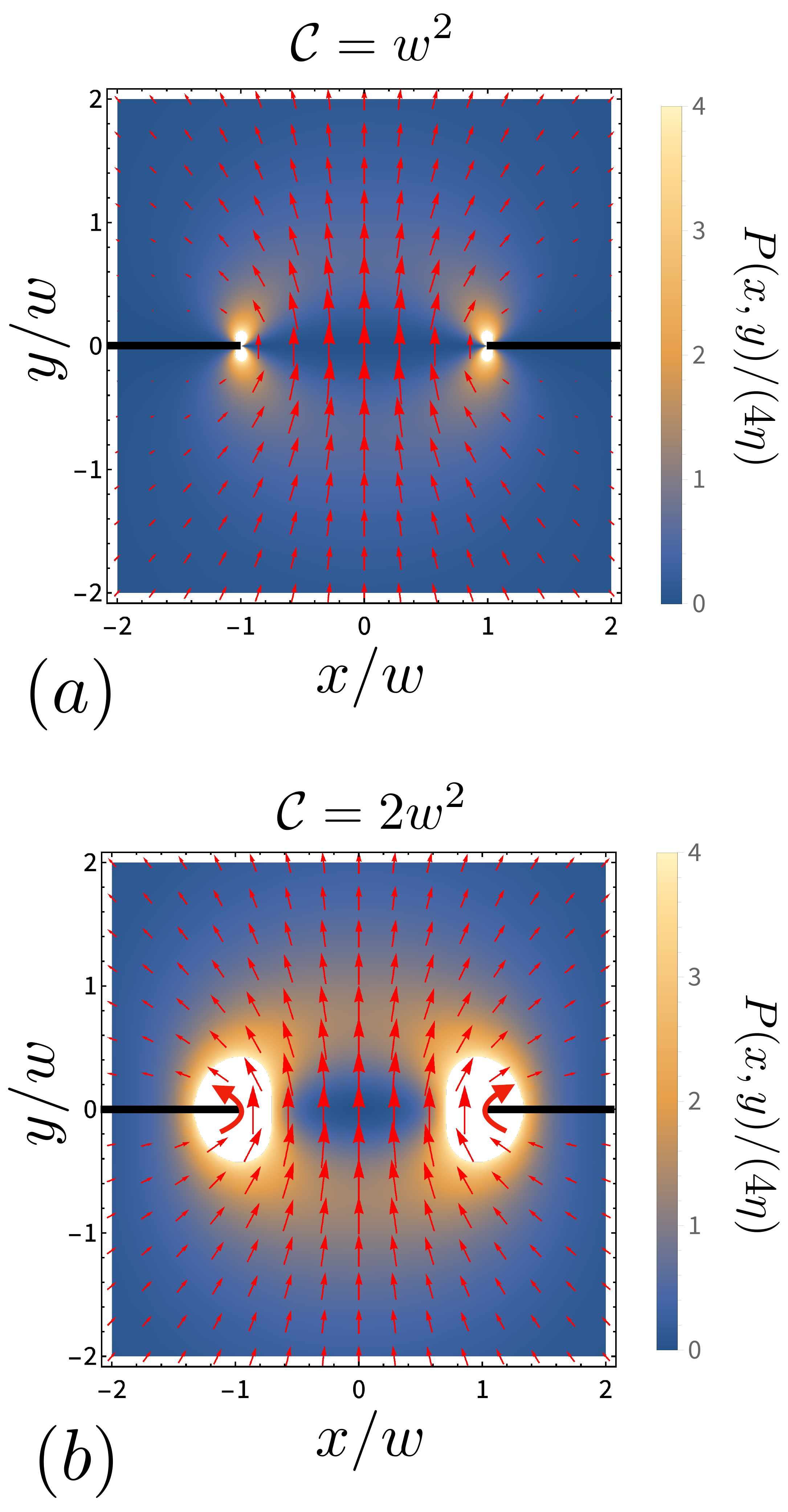}
  \caption{Velocity and dissipation profiles for (a) $\mathcal{C} = w^2$, and (b) $\mathcal{C} = 2 w^2$. The velocity directions and the magnitudes are shown by the direction and scale of red arrows, respectively. Note that the velocity streamlines are the same in both panels; the difference is only in the velocity magnitudes. In (b), with the choice $\mathcal{C} = 2 w^2$ (as in Ref.~\cite{PershogubaYongGlazmanPRB20}), the velocity diverges near the slit ends (as indicated by the thickest arrows), leading to a stronger divergence in the dissipation power ($1/\rho^3$ for $\mathcal{C} = 2 w^2$ vs.~$1/\rho$ for $\mathcal{C} = w^2$), cf. Eqs.~\eqref{eq:near_slit_power} and \eqref{P-1R}. As a result, the space-integrated   dissipation power, while being finite for $\mathcal{C} =w^2$, diverges for $\mathcal{C}=2w^2$. In fact, a similar dissipation profile, leading to the divergence of the total dissipation, is established for any $\mathcal{C} \neq w^2$. }
  \label{fig:velocity_dissipation}
\end{figure}

The divergence of dissipation for $\mathcal{C} \neq w^2$ is accompanied by a divergence of velocity near two slit endpoints, as shown in  Fig.~\ref{fig:velocity_dissipation}.
To see this explicitly, we evaluate the velocity near the right end of the slit using polar coordinates $\rho$ and $\varphi$ defined as
\begin{equation}
\label{polar-slit}
    y = \rho \sin\varphi,\quad x = w - \rho\cos\varphi.
\end{equation}
Here, $\varphi = 0$ in the slit (i.e., for $y= 0$ and $x < w$), and grows clockwisely.
When close enough to the slit right end ($\rho\ll w$), the velocity reads
\begin{equation}
\begin{aligned}
v_x & = \frac{eNV}{4\eta} \sqrt{2\rho w} \left( \cos^2\frac{\varphi}{2} \sin\frac{\varphi}{2} - \frac{w^2 -\mathcal{C}}{4w\rho} \sin\frac{\varphi}{2} \right)  \\
v_y & = \frac{eNV}{4\eta} \sqrt{2\rho w} \left(  \cos^3\frac{\varphi}{2} - \frac{w^2 - \mathcal{C}}{4w\rho} \cos\frac{\varphi}{2} \right).
\end{aligned}
\label{eq:near_slit_velocity}
\end{equation}
Equation~\eqref{eq:near_slit_velocity} is valid once $|\varphi| < 2\arccos\left[ (\rho/4w)^{1/4}\right]$, otherwise the next-leading order contribution is not negligible.
Clearly, when $\mathcal{C} \neq w^2$, velocities diverge as $v_x,v_y \sim \rho^{-1/2}$ near the slit endpoints.

From Eq.~\eqref{eq:near_slit_velocity} one can see  that the solution of the point-source model (i.e., of Ref.~\cite{FalkovichLevitovPRL17}) should be used with care. Indeed, for arbitrarily small $w$, we need to take care about distances $\rho$ much smaller than $w$, where a singular velocity profile is expected. Elimination of the singularity at such distances is a key argument in choosing $\mathcal C=w^2$ (see also discussion  in Appendix \ref{app:the_missing_solution}~\cite{Supmat}).

In agreement with Ref.~\cite{PershogubaYongGlazmanPRB20}, 
the value of $\mathcal{C}$ also changes the velocity profiles away from the slit. Indeed, with a general value of $\mathcal{C}$, velocities at large distances $|z| \gg w$ become
\begin{equation}
\begin{aligned}
v_x & \!\approx  v_r\! \left( \!\frac{x y^2}{r^3}\! -\! \frac{w^2 - \mathcal{C}}{2 w^2} \frac{x}{r}\! \right) \! = \!  v_r \!\left(\! \sin^2\!\varphi - \frac{w^2 - \mathcal{C}}{2 w^2}  \!\right)\!\cos\varphi, \\
\! v_y &\! \approx\!  v_r\! \left(\! \frac{y^3}{r^3}\! -\! \frac{w^2 - \mathcal{C}}{2w^2} \frac{y}{r}\! \right)\! = \! v_r \!\left(\! \sin^2\!\varphi - \frac{w^2 - \mathcal{C}}{2 w^2}  \!\right)\!\sin\varphi,
\end{aligned}
\label{eq:large_r_v}
\end{equation}
where $v_r = eNV w^2/\left(4r\eta\right)$.

Noticeably, when $\mathcal{C} = w^2$, following  Eq.~\eqref{eq:umain}, $v_x (x,0) = \partial_y v_x (x,y)|_{y \to 0}=0$ at the boundary, indicating that both the no-slip and no-stress boundary conditions are satisfied simultaneously. As seen from asymptotic Eq.~\eqref{eq:large_r_v}, this happens because  $v_x$ tends  to zero very quickly, $\propto y^2,$ when approaching to the boundary.    

It is worth noting, that  the finite-dissipation requirement should be undoubtedly satisfied for all geometries. However, in  most geometries the no-stress and no-slip boundary conditions lead to different solutions with finite but different total dissipations.
The simplest examples are the Poiseuille flow in the strip with fixed velocity in the middle of the strip or a flow around a disk with a fixed velocity at infinity. 
Specific property of the current  problem  is a  mentioned above
dependence $v_x \propto y^2, $
 which guarantees   that both  $v_x$ and $\partial_y v_x $ tend to zero for $y \to 0 $.

Importantly, the slip length entering   partial-slip boundary condition, Eq.~\eqref{eq:boundary_condition}, turns out to be   irrelevant and does not enter into the solution for velocity.  Indeed, when both the velocity and the shear stress are equal identically to zero, the equality of Eq.~\eqref{eq:boundary_condition} holds for an arbitrary value of $\lambda$. We see that long-distance behavior of the velocity flow is non-singular for any $\mathcal C.$  The  cases $\mathcal C=w^2$ and $\mathcal C=2 w^2$ 
 differ only in slightly different angular dependencies.
However, it is worth stressing that the large-distance behavior given by Eq.~\eqref{eq:large_r_v}  for $\mathcal C=w^2$ does not coincide with the behavior obtained in Ref.~\cite{FalkovichLevitovPRL17} within the ``point-source'' model; see discussion in Appendix 
\ref{app:the_missing_solution}.

%%%%%%%%%%%%%%%%%%%%%%%%%%%%%%%%%%%%%%%%%%
%%%%%%%%%%%%%%%%%%%%%%%%%%%%%%%%%%%%%%%%%%

Let us  discuss the energy balance for the case of $\mathcal{C} = w^2$. The dissipation power [in polar coordinates $(r,\theta)$ centered at $x=y=0$ with $\theta=0$ corresponding to $x>0$, $y=0^+$] reads
\begin{equation}
\begin{aligned}
P(r,\theta)  = \frac{ \left( e N V w^2 \right)^2}{16\eta} \frac{4 r^2 \sin^2(\theta)}{[r^4 + w^4 - 2 w^2 r^2 \cos(2\theta)]^{3/2}}.
\end{aligned}
\label{eq:full_dissipation}
\end{equation}
Near the slit endpoints it   becomes
\begin{equation}
P\propto  1/\sqrt{(x\pm w)^2 + y^2} \sim 1/\rho 
\label{P-1R}
\end{equation} and is thus integrable. 
The total dissipated power 
for the solution with $\mathcal{C}=w^2$ reads
\begin{equation}
  \mathcal{P} =  \frac{\pi}{8} \frac{(e N V w)^2}{\eta}.
    \label{eq:superballistics_dissipation}
\end{equation}

From energy conservation, the dissipated power equals to the energy supply $V^2 /R$
(Joule law). We thus obtain the resistance from viscosity-induced dissipation:
\begin{equation}
R = \frac{8}{\pi} \frac{\eta}{e^2 N^2 w^2}.
\label{eq:viscous_resistance}
\end{equation}
We can compare this result with the direct calculation of resistance. Inside the slit, the solution \eqref{eq:umain} with $\mathcal{C}=w^2$ yields 
$$v_y(x,y=0) = \frac{eNV}{4 \eta} \sqrt{w^2-x^2}, \quad |x|<w.$$ 
Then, for the total current, we have
\begin{equation}
   I = eN \int_{-w}^{w} v_y(x, 0) dx 
   = \frac{(eNw)^2 V \pi}{8 \eta}.
\end{equation}
By the direct definition of the resistance, we find $$R = \frac{V}{I}=\frac{8\eta}{\pi e^2 N^2 w^2}=R_\text{no-slip},$$ 
where $R_\text{no-slip}$
was obtained in Ref.~\cite{GuoPNAS17} for the no-slip boundary condition.
This result coincides with the result obtained from the dissipation, Eq.~\eqref{eq:viscous_resistance},
indicating  consistency of our analysis.

 Let us discuss this point in more detail. There exist at least three different methods to obtain the viscosity-induced resistance: (i) via evaluating the total viscosity-induced dissipation power
$R_P=V^2/\mathcal{P}$; (ii) via the definition of the resistance, i.e., $R = V/I$, where $I$ is the total current through the slit; and (iii) via evaluating the total work $W$ done by the electric force: $R_W=V^2/W$.
Above we have proved the agreement between resistance obtained from methods (i) and (ii). We will now show that the total work done by the electric force equals to the dissipation caused by viscosity.

The local work done by the electric field ($\mathbf{E}$) 
can be found from the Stokes equation:
\begin{equation}
\begin{aligned}
Ne \mathbf{v} \cdot \mathbf{E} & = - N e \mathbf{v}  \cdot \nabla \phi =- \eta \mathbf{v} \cdot \nabla^2 \mathbf{v}\\
& = -\eta \left[ v_x \left( \partial_x^2 + \partial_y^2 \right) v_x + v_y \left( \partial_x^2 + \partial_y^2 \right) v_y \right] \\
& = -2\eta \left[ u \frac{\partial^2}{\partial z \partial \bar{z}} \bar{u} + \bar{u} \frac{\partial^2}{\partial z \partial \bar{z}} u \right].
\end{aligned}
\label{eq:local_electric_work}
\end{equation}
The total work done by the electric force is equal to the integral of Eq.~\eqref{eq:local_electric_work} over space, i.e.,
\begin{equation}
\begin{aligned}
W & = Ne \iint dx dy \ \mathbf{v} \cdot \mathbf{E} \\
& = -4\eta \! \int \! d\bar{z} u\partial_{\bar{z}} \bar{u} \Big|_{\text{boundary }} \! -  \!4\eta \! \int \! dz \bar{u}\partial_{z} u \Big|_{\text{boundary}}\\
& +4\eta  \iint dx dy\  \frac{\partial u}{\partial z} \frac{\partial \bar{u}}{\partial \bar{z}},
\end{aligned}
\label{eq:work_before_int}
\end{equation}
where the second line refers to boundary contributions.
Following Eq.~\eqref{eq:large_r_v} for the velocity profile away from the slit, the work of the electric field at infinity is negligible.
Further, we can see from our solution \eqref{eq:umain} with $\mathcal{C}= w^2$ that the contribution to the total work from the boundary of the barrier also equals zero. 
The last term in Eq.~(\ref{eq:work_before_int}) exactly equals the total dissipation power produced by viscosity, Eqs.~\eqref{eq:Pzz} and \eqref{eq:superballistics_dissipation}, $ W=\mathcal{P},$
hence, $R_W=R_P.$
We have thus proven that the resistance evaluated by all three methods is the same.

In contrast, these three quantities are  different for the velocity profile obtained in Refs.~\cite{FalkovichLevitovPRL17,PershogubaYongGlazmanPRB20} for the no-stress boundary condition with Eq.~(\ref{ref_C}).
To start, Eq.~\eqref{eq:near_slit_power} indicates that at $\mathcal{C} = 2w^2$, the dissipation power near the end of the slit has a singularity that generates a divergent total dissipation.
The viscous resistance $R$ obtained from method (i) introduced above (i.e., $R_P=V^2/\mathcal{P}$) vanishes, because of the divergence of total dissipation $\mathcal{P}$.
Interestingly, if one fixes the total current $I$ rather than the voltage $V$ (i.e., if resistance is defined as $\mathcal{P}/I^2$), it would diverge for $\mathcal{C} = 2w^2$.
This discrepancy indicates that the choice $\mathcal{C} = 2w^2$ (as well as any other choice $\mathcal{C}\neq w^2$) is physically unreasonable, although it satisfies the Stokes equation with the no-stress boundary conditions,   and gives finite current through the slit. 
At the same time, resistances $R=V/I$ and $R_W$ obtained for $\mathcal{C}=2w^2$ from methods (ii) and (iii), respectively, are both finite, but are not equal to each other.
Indeed, Ref.~\cite{PershogubaYongGlazmanPRB20} reports a finite viscous resistance $R_\text{no-stress} = R_\text{no-slip}/2$ following method (ii), i.e., with the resistance defined as $V/I$. 
The resistance $R_W$, however, is equal to $2 R_\text{no-slip}/3$, when it is obtained by method (iii) from the total work of the electric field ($R = V^2/W$) in the bulk of the sample (see Appendix~\ref{app:extra_electric_work}). 
 
Let us note that the contribution to velocity of terms proportional to $(\mathcal{C}-w^2)$ [e.g., the second terms of Eqs.~\eqref{eq:near_slit_velocity}, \eqref{eq:large_r_v}]
does not produce any electric field in the sample, as
follows from Eq.~\eqref{eq:stokes_eq}.
Indeed, the whole family of solutions Eq.~\eqref{eq:umain} is characterized by the same electric potential. Thus, the situation with 
the velocity terms  $\propto (\mathcal{C}-w^2)$ is similar to the one encountered in Ref.~\cite{ShavitShytovFalkovichPRL19}, where the velocity profile is characterized by the coexistence of zero electric field and finite viscous dissipation in the bulk of the sample. In that paper, the energy conservation was fixed by the presence of a potential jump (infinite electric field) at the interface between the sample and metallic contact. It is clear, however, that in a realistic setup with metallic contacts in the slit geometry, the divergent (for $\mathcal{C}\neq w^2$) total viscous dissipation power cannot be balanced by any finite potential jump at the interfaces with metallic contacts.

\section{Analysis of the problem at arbitrary slip length}
\label{sec:finite_lambda}

Interestingly, although the solution in Eq.~\eqref{eq:umain} is obtained under the no-stress boundary condition, for $\mathcal{C}=w^2$ it also solves the general flow problem at arbitrary $\lambda$, including the no-slip limit as a particular case. However, there is a question about the uniqueness of this solution in the general case of arbitrary slip length $\lambda$, which cannot be answered without a general solution. 

Assuming that the total current through the slit is finite, the velocity is integrable at the $y=0$ boundary and goes to zero at $y\to+\infty$, we can express the velocity profile $\vec{v}(x,y) = (v_x(x, y), v_y(x, y))$ in the upper half-plane through $v_x(x, 0)$ and $v_y(x, 0)$: 
\begin{equation}
\begin{split}
\vec{v}(x, y) = \int \vec{K}_1(x-x', y) v_{x}(x', 0) dx'  \\ + \int \vec{K}_2(x-x', y)  & v_{y}(x', 0) dx'
\end{split}
\label{eq: convolution}
\end{equation}
with kernels
\begin{equation}
\begin{aligned}
& \vec{K}_1(x-x',y) =
\frac{2(x-x')y}{\pi [(x-x')^2+y^2]^2}
\left(\begin{array}{ll} x&\!\!\!-x'\\ &\!\!y\end{array}\right), \\
& \vec{K}_2(x-x',y) =
\frac{2y^2}{\pi [(x-x')^2+y^2]^2}
\left(\begin{array}{ll} x&\!\!\!-x'\\ &\!\!y\end{array}\right).
\end{aligned}
\label{eq:Kernels}
\end{equation}
Derivation of these kernels is presented in Appendix~ \ref{app:derivation of the Kernels}.

The boundary velocities should be found by imposing the boundary conditions. The first condition is given by Eq.~\eqref{eq:boundary_condition} and the second condition comes from the requirement of symmetry along the $y=0$ axis: 
$$\left.\partial_y^2 v_x(x,y)\right|_{y=0,|x|<w}=0.$$
\newline
It is convenient that the boundary velocities are defined on disjoint sets: (i) as the wall is impenetrable, we have $v_y(x,0)\equiv 0$ for $|x|>w$ and (ii)  from the symmetry of the problem, we have $v_x(x,0)\equiv 0$ for $|x|<w$. 
After imposing these boundary conditions, we get the following integral equations:
\begin{equation}
\begin{aligned}
v_x(x, 0) & = \frac{4 \lambda}{\pi} \frac{\partial}{\partial x} \fint_{w}^{\infty} \frac{x'}{x'^2-x^2}v_x(x',0) dx',
\\
& 0=\frac{\partial^2}{\partial x^2} \fint_{-w}^{w} \frac{1}{x'-x} v_y(x',0)dx'.
\end{aligned}
\label{eq:integral equations}
\end{equation}
Integrating both parts of the first equation over $x$, we get
\begin{equation}
    \int_{w}^{\infty} v_x(x', 0) dx' = - \frac{4 \lambda}{\pi} \int_{w}^{\infty} \frac{x'}{x'^2-w^2}v_x(x', 0)  dx'.
\end{equation}

If we are looking for the solution, for which $v_x(x, 0)$ does not change its sign on the wall (this is a reasonable assumption for a viscous flow without turbulence), the only possible solution is $v_x(x,0)=0$. Actually, this can be rigorously proved assuming integrable dissipation near the slit end points, see Appendix~\ref{strict proof for velocity}.
Thanks to the symmetry $v_y(x, y)=v_y(-x,y)$, the second equation is equivalent to 
\begin{equation}
     \fint_{-w}^{w} \frac{1}{x'-x} v_y(x',0)dx' = -\pi c_1 x.
    \label{rewriting}
\end{equation}

This is a singular integral equation, for which the general solution is known and can be found, e.g., in Ref.~\cite{Inteq}. This solution allows for the square-root and inverse square-root dependencies on $1-x^2/w^2$:
\begin{equation}
v_x(x,0)=0,\quad v_y(x,0)= c_1\sqrt{w^2-x^2} + \frac{c_2}{\sqrt{w^2-x^2}}.
\label{eq: general boundary v}
\end{equation}

The no-stress $\lambda \rightarrow \infty$ limit is singular. In this limit, the first line of Eq.~\eqref{eq:integral equations} 
reduces to 
\begin{equation}
    \frac{\partial}{\partial x} \fint_{w}^{\infty} \frac{x'}{x'^2-x^2}v_x(x',0) dx' = 0.
\label{eq:inteq no stress case}
\end{equation}
This equation actually allows nonzero solutions with velocity which does not change sign on the wall. From Eq.~\eqref{eq:better inteq for v_x}, we find, that the corresponding family of solutions for $v_x(x, 0)$ is \cite{Inteq}
\begin{equation}
    v_x(|x|>w, y\to 0) = c_3\frac{\mathrm{sign} (x) ~\mathrm{sign}(y)}{\sqrt{x^2-w^2}}
\label{eq:v_x with infinite lambda}
\end{equation}
and $v_y(|x|<w, 0)=0.$ 
Notice, that this family of solutions is larger than obtained for the no-stress case in Sec. \ref{sec:model}, Eq.~ \eqref{eq:umain}.
Indeed, computing the velocity in the whole plane, we find for the general no-stress solution:
\begin{equation}
\label{general_noslip}
u(z, \bar{z}) \equiv v_y + i v_x = \sum_{k=1}^{3} c_k U_k(z,\bar{z}),
\end{equation}
where  
\begin{equation}
\label{general_noslipU1}
U_1(z,\bar{z})=\frac{1}{2i}\left(\sqrt{\bar{z}^2-w^2}+\frac{z \bar{z}-w^2}{\sqrt{z^2-w^2}}\right),
\end{equation}
\begin{equation}
\label{general_noslipU2}
U_2(z,\bar{z})=\frac{i}{2} \left(\frac{2 z^2-z\bar{z}-w^2}{(z^2-w^2)^{3/2}}+\frac{1}{\sqrt{\bar{z}^2-w^2}}\right),
\end{equation} 
and 
\begin{equation}
\label{general_noslipU3}
U_3(z,\bar{z})=-\frac{i}{2} \left(\frac{w^2-z\bar{z}}{\left(z^2-w^2\right)^{3/2}}+\frac{1}{\sqrt{\bar{z}^2-w^2}}\right)
\end{equation}
[velocity profiles in terms of real coordinates $(x,y)$ can be found in the    Appendix \ref{even_more_singular_app}].

We thus see that the velocity field delivered by Eq. \eqref{eq:umain} is given by $c_1 \mathbf u_1+ c_2 \mathbf u_2+ c_3 \mathbf u_3 $, where $\mathbf u_k= (u_{kx},u_{ky}),$ $u_{kx}$, and $u_{ky}$ are real and defined by  equality $u_{ky} + i u_{kx}=U_k$, 
\begin{equation}
c_1=\frac{eNV}{4\eta },\text{ and }c_2=c_3=\frac{eNV}{8\eta}(\mathcal{C}-w^2).
\label{ci-for-Glazman}
\end{equation}
 We notice that the voltage drop across the sample, 
 $ V = V(y=\infty)-V(y=-\infty)$, corresponding to the solution Eq.~\eqref{general_noslip} depends on the coefficient  $c_1$ only:  $ V = ({4\eta}/{eN})c_1$.
This happens because electrical potentials corresponding to solutions  $\mathbf u_2$ and $\mathbf u_3$   become zero at infinity (see also discussion in Appendix \ref{even_more_singular_app})  

Finally, computing vorticity,
\begin{equation}
    \omega(z, \bar{z}) = 2\mathrm{Im} \Big{[} c_1\frac{z}{\sqrt{z^2-w^2}}+ (c_2-c_3)\frac{z}{(z^2-w^2)^{3/2}}\Big{]},
    \label{eq:general_vorticity}
\end{equation}
we find that the solutions with $c_2 \neq c_3$ have non-integrable divergence of vorticity, and, therefore, do not follow the Keldysh-Sedov theorem.

Inspecting the velocity profiles of Eq. \eqref{general_noslip}, we find that the solution with finite dissipation is unique and is identical to the one delivered by  Eq.~\eqref{eq:umain} at $\mathcal{C} = w^2$. This solution in fact satisfies both the no-slip and no-stress boundary conditions.
As discussed in Appendix~\ref{app:the_missing_solution}, this solution cannot be captured by using the partial-slip ``point-source'' solution 
with $w\equiv 0$ (Ref.~\cite{FalkovichLevitovPRL17}).

\section{Temperature Profiles}
\label{sec:temperature}

With the velocity profiles of Eq.~\eqref{eq:umain}, in this section, we derive the corresponding temperature distribution.  
We start with  noting that  the electronic heat transfer in the  hydrodynamic   regime  strongly depends on the spectrum.  
In particular, for  relativistic dispersion, the electron-electron scattering does not affect the energy current, as it is represented through the total momentum conserved by scatterings \cite{NarozhnyAofPhys19}. On the contrary, for the parabolic dispersion, electron-electron scattering is capable of establishing a finite thermal conductivity without extra momentum-relaxing scattering events. Here, we limit ourselves with discussion  of the parabolic spectrum only. Heat transfer in the system of Dirac fermions will be discussed elsewhere.  

 For parabolic dispersion, the temperature obeys the heat-balance equation \cite{Landafshiz}
\begin{equation}
    -\kappa \nabla^2 \delta T + C (\mathbf{v} \cdot \mathbf{\nabla} )\delta T =  P(x,y) - N\gamma \delta T.
    \label{eq:heat_balance}
\end{equation}
Here, $\delta T \equiv T - T_0$ refers to the temperature variation with respect to the background temperature $T_0$, $\kappa$ is the heat conductivity, $C$ the heat capacity, and $\gamma$ denotes the electron-phonon coupling constant. 
It is worth noting that  $\kappa,~C,$ and $\gamma$  depend on temperature so that there is some back reaction.  Here, we linearize heat balance equation, assuming that $\delta T \ll T_0.$    The conditions for validity of linearization are different for the cases of small and large heat conductivity [$T_V \ll T_0$ and $T_{\rm scale} \ll T_0,$   respectively, where $T_V$  and   $T_{\rm scale}$    are characteristic  overheating  scales for these cases determined below by Eqs.~\eqref{scales} and \eqref{scale1}].

The ``convection'' term, 
${\rm div}(\mathbf v T)=(\mathbf{v} \cdot \mathbf{\nabla} )\delta T$, which breaks the particle-hole symmetry, is known to introduce an asymmetry in the temperature distribution~\cite{Tikhonov2019,GuPRB21}.
Throughout this section, we consider the regular (no-stress--no-slip) velocity profile, i.e., Eq.~(\ref{eq:umain}) with $\mathcal{C}=w^2$, which is the unique choice that produces a finite total dissipation, as discussed above.

Generally, it is difficult to obtain the full solution of Eq.~\eqref{eq:heat_balance}.
Here, for simplicity, we focus on the two limiting cases of small and large heat conductivity.

\subsection{Small heat conductivity}
\label{sec:small_kappa}
In this section, we ignore the heat conductivity (i.e., we take $\kappa = 0$).
Then, equation \eqref{eq:heat_balance} reduces to 
\begin{equation}
    C \left(v_x \frac{\partial \delta T}{\partial x} + v_y \frac{\partial \delta T}{\partial y} \right) = P(x,y) - N \gamma \delta T
    \label{eq:first case}
\end{equation}
Using the method of characteristics and switching to the complex coordinates, we find: 
\begin{equation}
    \begin{cases}
         \ C\dfrac{d \delta T}{d t} =  P(z, \bar{z}) - N\gamma \delta{T}, \\[0.5cm]
        \ \dfrac{dz}{dt} = i\bar{u},\quad \dfrac{d\bar{z}}{dt} = -iu.
    \end{cases}
    \label{eq:solvation_system}
\end{equation}
Here, the source term $P(z, \bar{z})$ is defined by Eq.~\eqref{eq:Pzz} and $t$ counts the ``time'' passed since the movement along the characteristic curve started.

Solutions of characteristic differential equations consist of a set of hyperbolas parametrized by $r_0$,
\begin{equation}
    \frac{x^2}{r^2_0} - \frac{y^2}{w^2-r_0^2} = 1,
\end{equation}
or, equivalently,
\begin{equation}
z \bar{z} + \sqrt{z^2-w^2}  \sqrt{\bar{z}^2 - w^2} = 2r^2_0-w^2.  
\label{eq:characteristics}
\end{equation}
Geometrically, $r_0$ refers to the minimum distance from point (0, 0) to a given characteristic curve. The following equation establishes the parameter $r_0$ of the characteristics passing through a given point $(r,\,\theta)$:
\begin{equation}
r_0^2 = \frac{1}{2} \left({ w^2 + r^2 - \sqrt{r^4+w^4 - 2r^2 w^2 \cos{2 \theta}}}\right). 
\end{equation}
The bounded solution along the characteristic satisfying $\delta T(-\infty)=0$ can be found as follows:
\begin{equation}
    \delta T(t) = \int\limits_{-\infty}^t
    \frac{P(s)}{C}\,\exp\left[-\frac{N\gamma(t-s)}{C}\right]\,ds .
    \label{eq:solution-0}
\end{equation}

\begin{figure}
  \includegraphics[width= 1.0\linewidth]{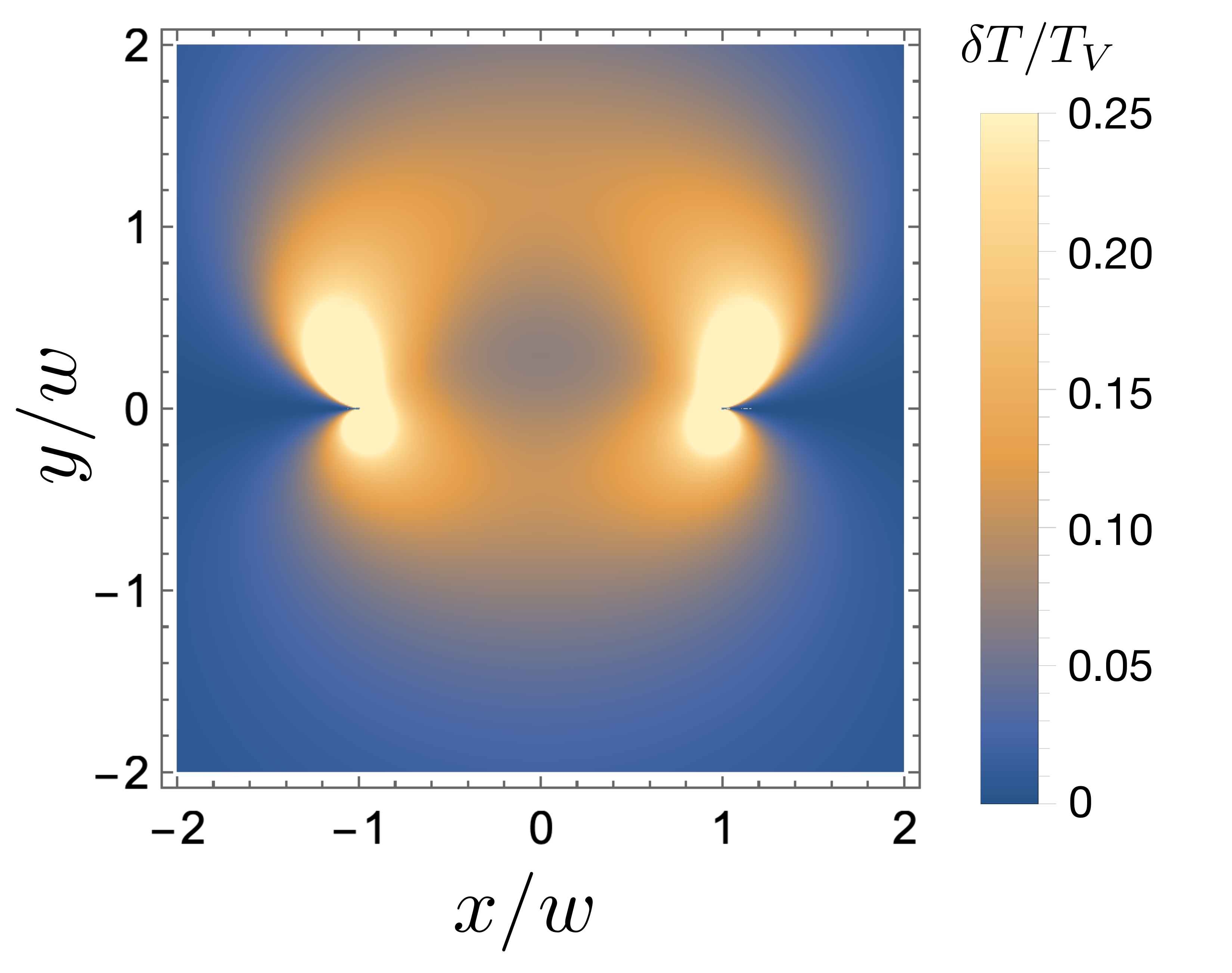}
  \caption{Heatmap of the nonequilibrium variation of the temperature $\delta T$ [defined in Eq.~\eqref{dTr}], computed according to Eq.~\eqref{eq:solution} for the velocity profile with $\mathcal{C}=w^2$ and  $l_\text{vis}=2w$. The temperature is measured in units of $T_V$, see Eq.~\eqref{scales}. Because of the divergence of $\delta T$ near the slit endpoints, the values are clipped at $\delta T = T_V/4$ for better visibility. The temperature profile, compared to the dissipation power shown in Fig.~\ref{fig:velocity_dissipation}(a), shows the $y\to -y$ asymmetry induced by the nonlinear character of the convection term in the heat-balance equation \cite{RokniLevinson1995,Tikhonov2019,GuPRB21}.}
  \label{fig:heatmap}
\end{figure}

Using Eq.~\eqref{eq:characteristics}, we can rewrite the velocity and dissipation profiles as follows:
\begin{equation}
    u(z, \bar{z}) = \frac{eNV}{4 \eta i} \frac{r_0^2-w^2}{\sqrt{z^2-w^2}},
    \label{eq:complex-velocity_new}
\end{equation}
\begin{equation}
    P(z, \bar{z}) = \frac{(eNVw)^2}{4 \eta} \frac{(w^2-r_0^2)(z \bar{z}-r^2_0)}{(z\bar{z}+w^2-2r^2_0)^3}.
    \label{dissipation_new}
\end{equation}
The dissipation pattern described by Eq.~(\ref{dissipation_new}) is shown in Fig.~\ref{fig:velocity_dissipation}(a).
It is very convenient that along a given characteristic ($r_0=\mathrm{const}$), dissipation depends only on $r^2=z\bar{z}$. Using Eqs.~\eqref{eq:solvation_system} and \eqref{eq:complex-velocity_new}, we get  the following differential equation for $r^2(t)$:
\begin{equation}
    \frac{dr^2}{d t} = \pm {\frac{eNV(w^2-r^2_0)}{4 \eta}} {\frac{\sqrt{(2r^2+w^2-2r^2_0)^2-w^4}}{r^2+w^2-2r^2_0}},
    \label{equation for r^2}
\end{equation}
with $+$ for the upper and $-$ for the lower half-planes, correspondingly. 
Solving this equation, we obtain the relation between the ``time'' on a characteristic curve and coordinates on the plane $r,\,r_0$:
\begin{equation}
    f(r^2) = {\frac{eNV}{ \eta}} \left|t-t_0\right|,
    \label{kinda_solution}
\end{equation}
where $t_0$ corresponds to the ``time'' in the slit ($y=0$) and $f(r^2)$ is given by
\begin{equation}
\begin{split}
       f(r^2) = & \frac{{\sqrt{[2(r^2-r^2_0)+w^2]^2-w^4}}}{(w^2-r^2_0)}+ \\
       & + \frac{w^2-2r^2_0}{w^2-r^2_0}\,\text{arccosh}\left(1+2\frac{r^2-r^2_0}{w^2}\right).
\end{split}
\end{equation}
We can now switch the integration variable in the integral \eqref{eq:solution-0} from $ds$ to $u=dr^2$ to compute the temperature:
\begin{equation}
\delta T=\delta T(\mathbf r)= 
\left \{ \begin{array}{l}  \delta T_+(\mathbf r),\quad {\rm for}\quad y>0,
\\
\delta T_-(\mathbf r),\quad {\rm for}\quad y<0,
\end{array}
\right.
\label{dTr}
\end{equation}
where
\begin{equation}
 \delta T_{\pm}(r_0, r)  =  \frac{ \displaystyle  \delta T_0(r_0) \! \pm \! \int  \limits_{r_0^2}^{r^2}\!I_{r_0}\!(\mathrm{u}) e^{\pm \zeta f(\mathrm{u})} \! \frac{d\mathrm{u}}{w^2}}{e^{\pm\zeta f(r^2)}}
 \label{eq:solution}
\end{equation}
is expressed 
in terms of the temperature in the slit (i.e., at $y=0,\;-w<x<w$):
\begin{equation}
\delta T_0(r_0) = \int  \limits_{r^2_0}^{\infty}\!I_{r_0}\!(\mathrm{u}) e^{\zeta f(\mathrm{u})} \! \frac{d\mathrm{u}}{w^2}
\label{dT0}
\end{equation}
where
\begin{equation}
I_{r_0}(\mathrm{u})= T_V \frac{w^4 (\mathrm{u}-r^2_0)}{(\mathrm{u}+w^2-2r^2_0)^2
\sqrt{[2(\mathrm{u}-r^2_0)+w^2]^2-w^4}},
\end{equation}
and
\begin{equation}
    \zeta = \frac{w^2}{l_\text{vis}^2}, \quad
    l_\text{vis} = \sqrt{\frac{eVC}{\eta \gamma}} w,\quad T_V = \frac{eNV}{C}.
    \label{scales}
\end{equation}
The last two expressions here define the spatial length scale $l_\text{vis}$ of the heated area and the corresponding temperature scale $T_V$ governing the variation of temperature in the heated region. The temperature profile given by Eq.~\eqref{eq:solution} is illustrated in Fig.~\ref{fig:heatmap}.

Far away from the slit, at $r \gg w$, we find
\begin{equation}
    \delta T_{\pm} (r, \theta) =  T_V \frac{w^2}{2 r^2} g_{\pm}\left(\frac{2 r^2}{l^2_\text{vis} \sin^2{\theta}}\right),
\end{equation}
where $g_{\pm}(t)=t \exp\left(\mp t\right) \text{Ei}\left(\pm t\right) \mp 1$ and $\text{Ei}$ is the exponential integral function. 
In the limit of $r \gg l_\text{vis} \sin \theta$, to the leading order of $l_\text{vis} /r$, the temperature distribution  becomes symmetric between the half-planes:
\begin{equation}
\delta T_{\pm}(r,\theta) \approx  T_V \frac{w^2 l_\text{vis}^2}{4r^4} \sin^2\theta.
\label{eq:distant_temperature}
\end{equation}
In this limit, the heat produced by the viscous dissipation is balanced by the phonon-emission, i.e., $ \delta T \approx C P(x,y)/N \gamma$. In this case, the temperature $\delta T$ and the velocity squared $v^2$ share the same features. Indeed, both of them decay as $\sim 1/r^4$ in space, with the directional distribution $\sim \sin^2\theta$.

\begin{figure}
  \includegraphics[width= 0.8\linewidth]{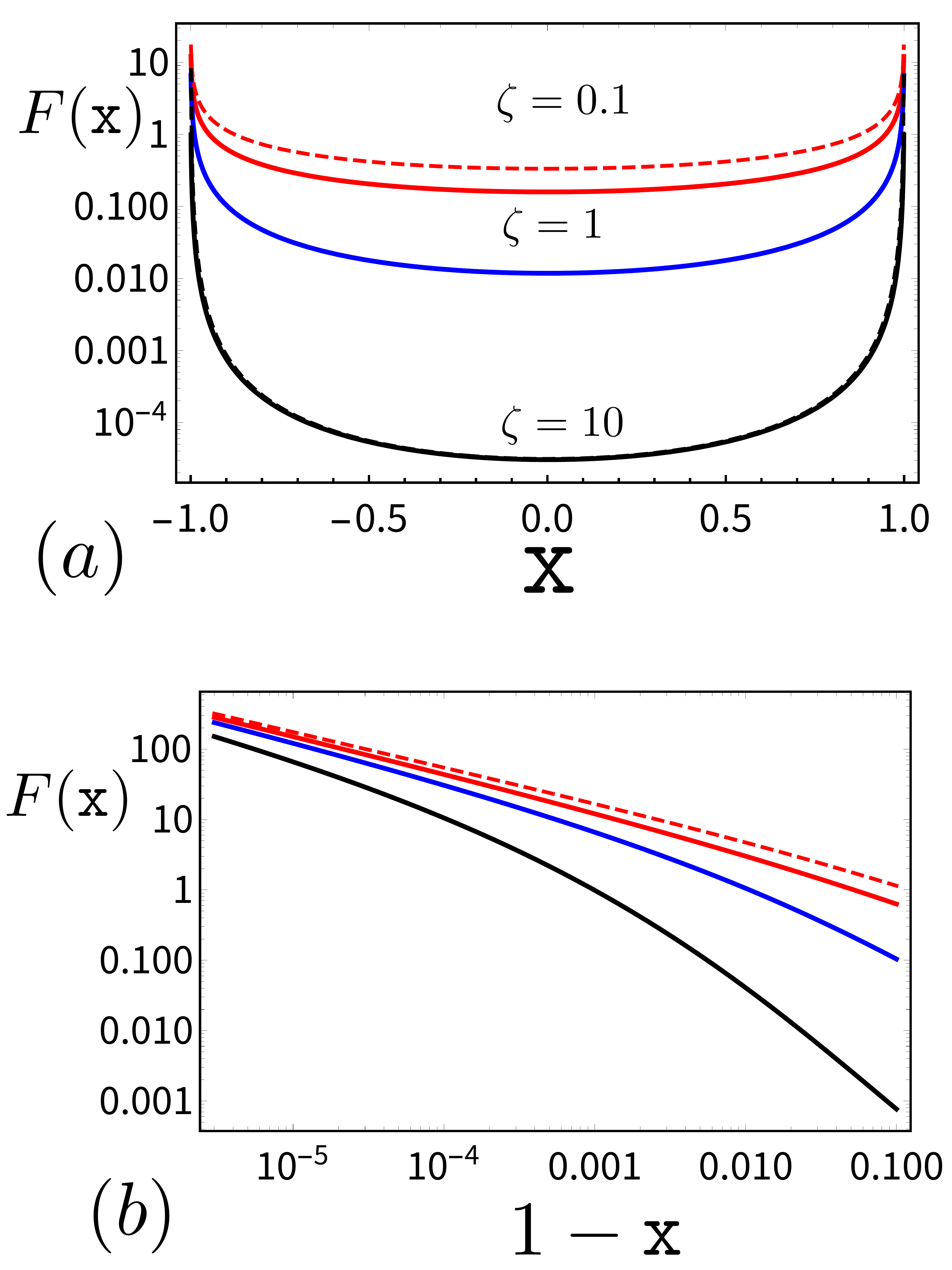}
  \caption{(a) Curves of the function $F(\texttt{x})$ [Eq.~\eqref{Fx}] in the slit (solid lines), for $\zeta = 10$ (black), $\zeta = 1$ (blue) and $\zeta = 0.1$ (red). The red and black dashed lines show asymptotic results described by Eqs.~\eqref{eq:zx} and \eqref{large-zeta}, respectively. The red lines agree better near the slit ends; The black ones instead have a better overlap in the middle of the slit (the curves are almost indistinguishable). (b) The asymptotic feature of corresponding curves in (a) near the right end of the slit (i.e., $\texttt{x} \to 1$). Near the slit endpoint, $F(\texttt{x})$ curves with different $\zeta$ values approach the same (red dashed) line described by Eq.~\eqref{eq:zx}.}
  \label{fig:fx}
\end{figure}

Let us now discuss the temperature distribution $\delta T_0(r_0=r)$ in the slit (for $|x|<w$, $y=0$), which can be found from Eq.~\eqref{dT0} as $$\delta T_0(r) = T_V F(\texttt{x}=r/w),$$ 
with   
\begin{align}
&F(\texttt{x})=\int_0^\infty \frac{ \mathrm{u} d\mathrm{u}}{2 \sqrt{\mathrm{u} (\mathrm{u}+1)} (\mathrm{u}+1-\texttt{x}^2)^2}
\label{Fx}
\\
&\!\times 
\exp\left\{\!\frac{-\zeta  \left[(1-2 \texttt{x}^2) \text{arccosh}(2 \mathrm{u}+1)+2 \sqrt{\mathrm{u} (\mathrm{u}+1)}\right]}{1-\texttt{x}^2}\right\}
\notag
\end{align}
a function of the dimensionless variable $\texttt{x} \equiv x/w$. We plot $F(\texttt{x})$ in Fig.~\ref{fig:fx}, for different $\zeta$ values.

%%%%%%%%%%%%%%%%%%%%%%%%%%%%%%%%%%%%%%%%%%
%%%%%%%%%%%%%%%%%%%%%%%%%%%%%%%%%%%%%%%%%%

In the limit $\zeta \sqrt{1-\texttt{x}^2} \gg 1$, the integral in Eq.~\eqref{Fx} is dominated by $u \ll 1$, resulting into
\begin{equation}
\label{large-zeta}
    F(\texttt{x})\approx \frac{1}{32\zeta^3(1-\texttt{x}^2)^2}.
    \end{equation}
In the opposite limit, $\zeta \sqrt{1-\texttt{x}^2} \ll 1$, the exponential factors in Eq.~\eqref{eq:solution} can be neglected. This gives for $F(\texttt{x})$ the following approximation:
\begin{equation}
    F(\texttt{x}) \approx \frac{1}{2 \texttt{x}^3}\left[\frac{\arcsin (\texttt{x})}{\sqrt{1-\texttt{x}^2}}-\texttt{x}\right].
    \label{eq:zx}
\end{equation}
This function has the following limiting behaviors: 
\begin{equation}
    F(0)=1/3, \qquad F(\texttt{x}\to 1)\approx \frac{\pi}{4\sqrt{2}}(1-\texttt{x})^{-1/2},
\end{equation} indicating a weak divergence of temperature upon approaching  the slit edge.

Let us consider the temperature distribution near the slit but away from the $y=0$ line, in the coordinates of 
Eq.~(\ref{polar-slit}).
In this case, the characteristics can be expressed as
\begin{equation}
\rho = \rho_0 \frac{1}{\cos^2\frac{\varphi}{2}},
\end{equation}
where $\rho$ are $\varphi$ refer to the polar-coordinate variables defined in Eq.~\eqref{eq:near_slit_velocity}. Each characteristic curve is labeled by its smallest distance $\rho_0$ to the right end of the slit.
We can now study the temperature distribution along each characteristic as a function of $\varphi$.
Indeed, following the introduced general method, the temperature near the endpoint of the slit is given approximately by
\begin{equation}
\begin{aligned}
&\mathcal{T} (\varphi)  = \mathcal{T}( \varphi_0) \exp \left\{-\frac{4w^2}{l^2_\text{vis}} \chi\, [ \xi(\varphi) - \xi (\varphi_0) ]\right\} \\
&\ + \! \frac{1}{4\chi} \! \int_{\varphi_0}^{\varphi} \!d\varphi'  \sin^2\!\frac{\varphi'}{2}  \exp\! \left\{ \frac{4w^2}{l^2_\text{vis}} \chi\, [ \xi(\varphi') - \xi (\varphi) ] \right\},
\end{aligned}
\label{eq:tl}
\end{equation}
where $\mathcal{T}\equiv \delta T/T_V$ is the dimensionless temperature,
\begin{equation}
    \chi \equiv\frac{N}{C} \sqrt{\frac{\rho_0 }{2w}},
    \label{eq:chi_def}
\end{equation}
 $\varphi_0$ is the starting point of the integral, and 
\begin{equation}
\xi (\varphi) = 2\tan\frac{\varphi}{2} \left( 1 + \frac{1}{3} \tan^2\frac{\varphi}{2} \right)
\end{equation}
is the angular distribution function.

When $\varphi_0 < \varphi$, the integral is along the transport direction of particles; otherwise, the integral is against the particle traveling direction. It is convenient to choose $\varphi_0 = 0$, since we know the temperature distribution inside the slit and near the edges ($\rho_0 \ll w/\zeta^2$) from Eq. \eqref{eq:zx}. Therefore,
\begin{equation}
    \mathcal{T}(\varphi_0) = \mathcal{T} (0) = T_V \frac{\pi}{8} \sqrt{\frac{2 w}{\rho_0}}.
\label{T0}
\end{equation}

%%%%%%%%%%%%%%%%%%%%%%%%%%%%%%%%%%%%%%%%%%
%%%%%%%%%%%%%%%%%%%%%%%%%%%%%%%%%%%%%%%%%%

%%%%%%%%%%%%%%%%%%%%%%%%%%%%%%%%%%%%%%%%%%
\begin{figure}
  \includegraphics[width=\linewidth]{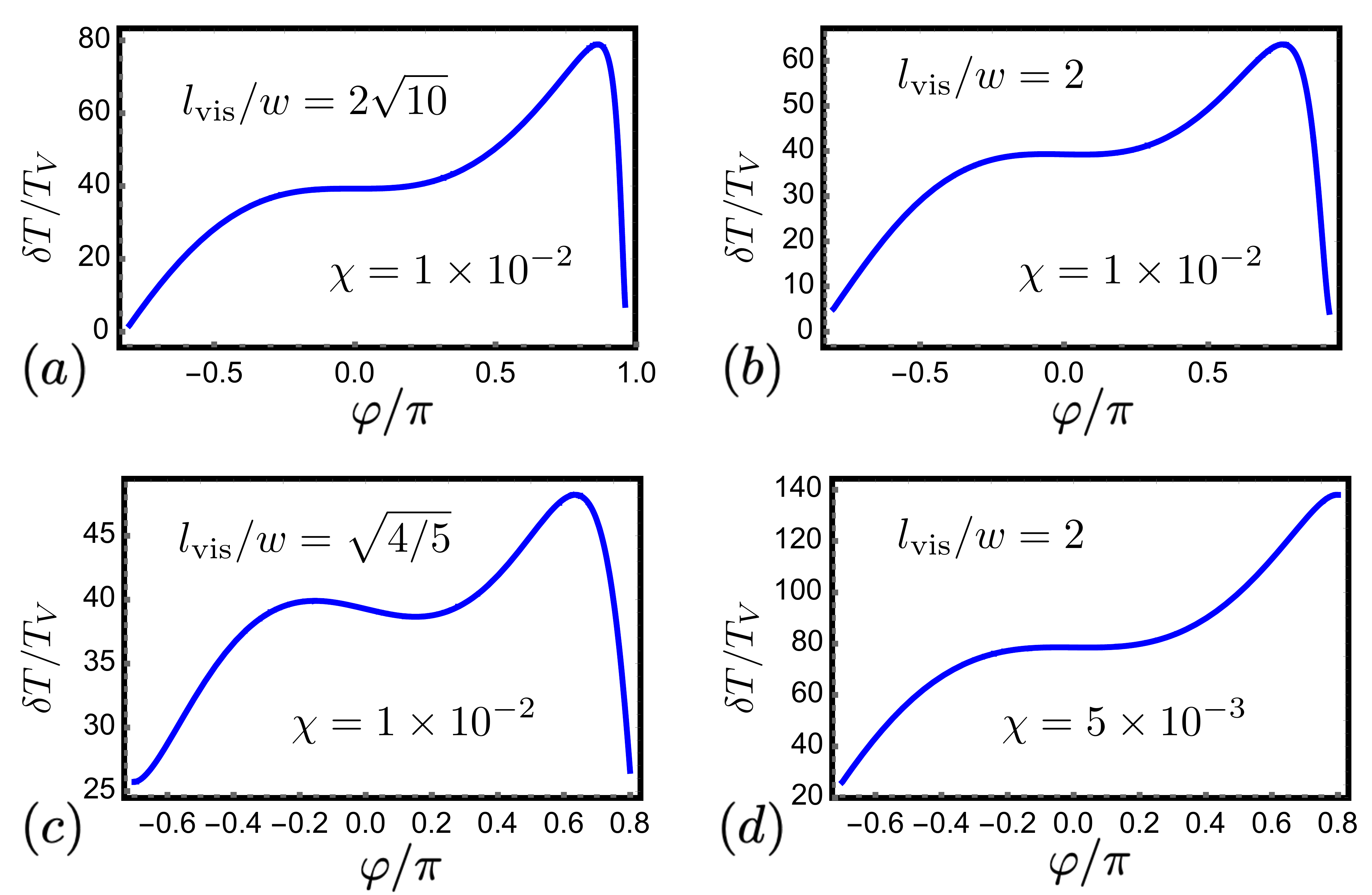}
  \caption{Temperature distribution (in units of $T_V$) near the right end of the slit, with different choices of $l_\text{vis}/w$ (different driving amplitude) and $\chi$ [different characteristics, see Eq.~\eqref{eq:chi_def}]. (a) $l_\text{vis}/w = 2\sqrt{10}$ and $\chi = 0.01$; (b) $l_\text{vis} /w = 2$ and $\chi = 0.01$; (c) $l_\text{vis} /w = \sqrt{4/5}$ and $\chi = 0.01$, and (d) $l_\text{vis} /w = 2$ and $\chi = 0.005$.}
  \label{fig:near_slit_temperature}
\end{figure}
%%%%%%%%%%%%%%%%%%%%%%%%%%%%%%%%%%%%%%%%%%

The most interesting result here is  the temperature divergence in the the vicinity of the slit endpoints: $T \propto \rho^{-1/2}$.
The divergence is cured  by taking into account finite heat conductivity.
For very small but finite heat conductivity, diffusion term in the heat balance equation can be neglected everywhere except narrow regions  near the endpoint of the slit.
Indeed, after neglecting the angle dependence, this  term  is estimated as  $\kappa\rho^{-2} \delta T$ near the slit endpoint. In comparison, since the velocity $\mathbf{v}\sim \sqrt{\rho}$ [see Eq.~\eqref{eq:near_slit_velocity}], the convection term is proportional to $\rho^{-1/2}$ and grows much slower than the heat diffusion contribution upon approaching the endpoint. Consequently,  the diffusion contribution dominates at positions close enough to the slit ends even for very small  $\kappa$.
By comparing two terms on the left side of Eq.~\eqref{eq:heat_balance}, one finds that the  diffusion comes into play  for
$$\rho < w\left(\frac{\kappa \eta}{eNCVw^2}\right)^{\frac{2}{3}}.$$

Figures~\ref{fig:near_slit_temperature}(a), (b), and (c) show the temperature profiles of the same characteristic curve $\chi = 0.01$. When $l_\text{vis}$ decreases (corresponding to a smaller driving), a two-peak structure of the temperature profile begins to emerge, and becomes apparent in Fig.~\ref{fig:near_slit_temperature}(c). In Fig.~\ref{fig:near_slit_temperature}(d), we take the same ratio $l_\text{vis}/w$ as that in Fig.~\ref{fig:near_slit_temperature}(b), but use a smaller value of $\chi = 0.005$. The corresponding shapes of the temperature profile of Figs.~\ref{fig:near_slit_temperature}(b) and (d) are rather similar. However, the temperature values in Fig.~\ref{fig:near_slit_temperature}(d) almost double, in agreement with the $\sim 1/\sqrt{\rho_0}$ dependence near the endpoint of the slit [see Eq.~\eqref{T0}].

\subsection{Large heat conductivity}
\label{sec:large_kappa}

Now, we move on to the opposite limit of a large heat conductivity. In this limit, neglecting the convection term, the heat-balance equation takes the form
\begin{equation}
    -\kappa \nabla^2 \delta T  =P(x,y) - N\gamma\delta T.
    \label{eq:large_kappa_balance}
\end{equation}
Equation~\eqref{eq:large_kappa_balance} can be solved by going into the momentum space, with the solution
\begin{equation}
    \delta T(x,y) = \frac{1}{2\pi\gamma}\! \iint dx' dy' \mathcal{K}(x -x',y-y') P(x',y'),
    \label{eq:large_kappa_temperature}
\end{equation}
where
\begin{equation}
\begin{aligned}
    \mathcal{K}(r) & =  \int_0^\infty k dk \frac{1}{1 + \frac{\kappa}{n \gamma} k^2} J_0(kr) \\
    & = \frac{1}{l_\text{diff}^2} K_0 \left( \frac{r}{l_\text{diff}} \right).
\end{aligned}
\label{eq:kernel}
\end{equation}
is the circularly symmetric kernel. 
In Eq.~\eqref{eq:kernel}, $J_0$ and $K_0$ are the modified Bessel functions, of the first and second kinds, respectively.

In Eq.~\eqref{eq:kernel}, we have defined another scale, 
$$l_\text{diff} \equiv \sqrt{\kappa/(N\gamma)},$$ 
that reflects the length scale of heat diffusion.
Indeed, the value of $l_\text{diff}$ increases in samples with either a strong heat conductivity $\kappa$, or a weak phonon emission rate $\gamma$.
When $\gamma$ decreases, heat diffuses to larger distances before being dissipated via phonon emissions.
In contrast to $l_\text{vis}$ that reflects non-equilibrium driving, $l_\text{diff}$ is a sample-specific quantity that is already defined at equilibrium.

Now we are in a position to evaluate the temperature profile with Eqs.~\eqref{eq:full_dissipation}, \eqref{eq:large_kappa_temperature}, and \eqref{eq:kernel}. The kernel $\mathcal{K}(r')$ decays exponentially $\propto \exp(-r'/l_\text{diff})$ when $r'$ becomes larger than $l_\text{diff}$.
Consequently, for positions far away from the slit, $r \gg l_\text{diff}$, the temperature profile in the large heat-conductivity limit approximately equals to Eq.~\eqref{eq:distant_temperature} for $\kappa = 0$, where temperature is proportional to the local dissipation power.
By contrast, for small distances, $r \ll l_\text{diff}$, the temperature profiles for the two limiting cases of large and small $\kappa$ are drastically different.

\begin{figure}
  \includegraphics[width=\linewidth]{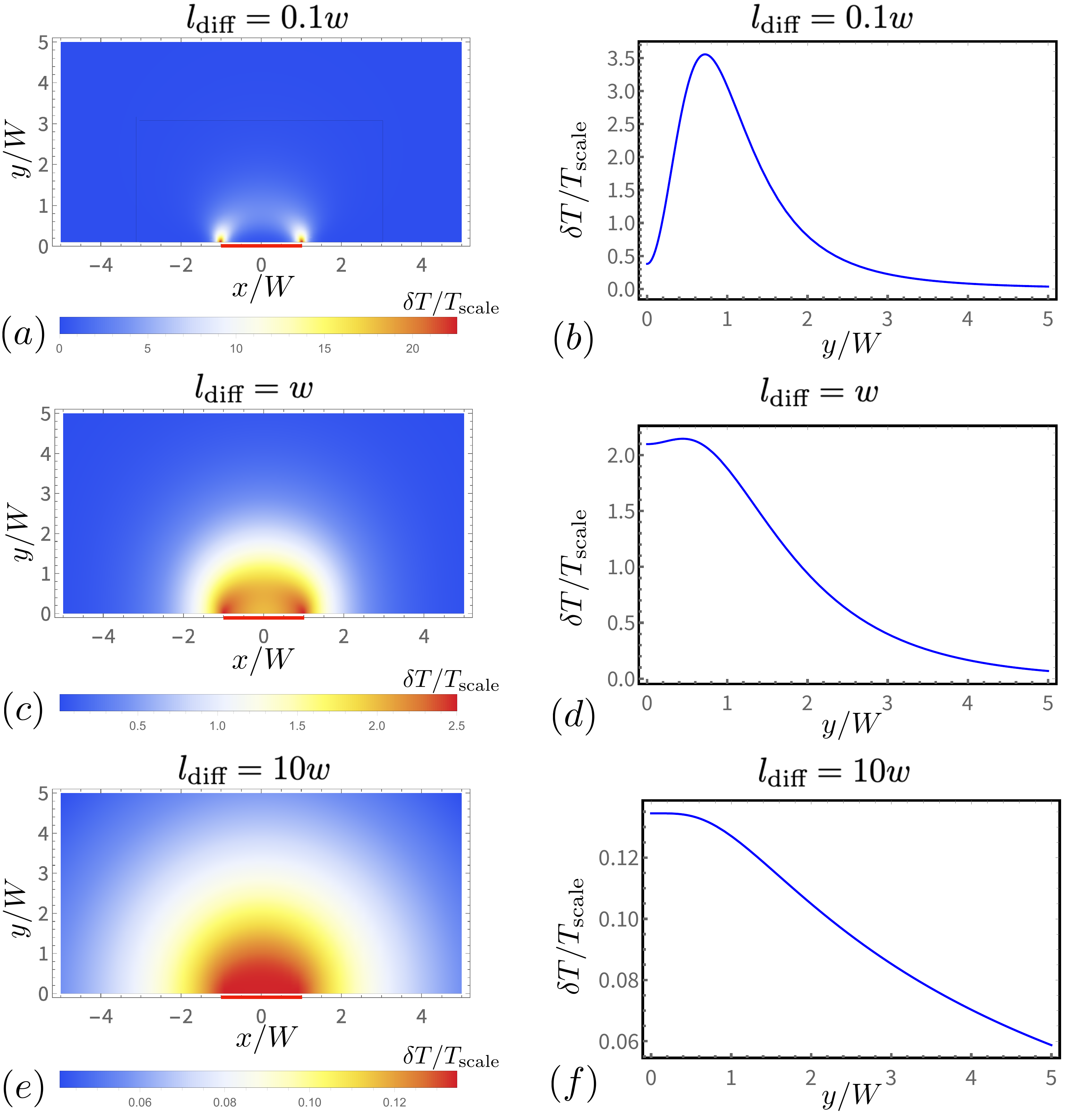}
  \caption{Temperature distribution [in units of $T_\text{scale} = e^2 N V^2/(32 \pi \gamma\eta)$] of the upper half-plane of a sample with a strong diffusion. We provide only temperature profile of the upper half-plane, since the distribution is symmetric (with respect to $y = 0$) without contribution from the convection term. The slit has been marked out by the red solid line. We choose $l_\text{diff}= 0.1 w$ for (a) and (b); $l_\text{diff} = w$ for (c) and (d); $l_\text{diff} = 10 w$ for (e) and (f). Figures of the right panel present temperature distribution at $x = 0$, when $y$ changes. The tail of an exponentially decaying temperature is clear in (b) and (d), but missing in (f).}
  \label{fig:diffusion_temperature}
\end{figure}

In the limit of a large heat conductivity, we work out the temperature profile via numerical integration of Eq.~\eqref{eq:large_kappa_temperature}. The results are presented in Fig.~\ref{fig:diffusion_temperature}, where \be 
T_\text{scale} = e^2 N V^2/(32 \pi \gamma\eta) 
\label{scale1}
\ee 
is another temperature scale.
Here we only show the temperature distribution of the upper half-plane $y > 0$ (since we  excluded the convection contribution, temperature distribution is symmetric with respect to the $y = 0$ line).
As seen from this figure,  heat conductivity  cures temperature singularities as expected.

%%%%%%%%%%%%%%%%%%%%%%%%%%%%%%%%%%%%%%%%%%
%%%%%%%%%%%%%%%%%%%%%%%%%%%%%%%%%%%%%%%%%%

The temperature profile is strongly influenced by the distance ratio $l_\text{diff}/w$.
When $l_\text{diff} \ll w$, most dissipation-induced heating is transferred into phonons before diffusion in space, and the area with hot electrons is strongly confined near two endpoints of the slit, as shown in Figs.~\ref{fig:diffusion_temperature}(a) and (b) for $l_\text{diff} = 0.1 w$. Here, the temperature profile shows a clear ring-shaped structure.
When $l_\text{diff}$ increases, the heated area begins to extend. When $l_\text{diff} = w$ [Figs.~\ref{fig:diffusion_temperature}(c) and (d)], the heated area is slightly larger than the slit size $w$, although the strong heating near slit endpoints remains clearly visible. Finally, when $l_\text{diff} = 10 w$, phonon emission is negligible at the scale of the slit width $w$, and the area with hot electrons becomes extensive.
In Figs.~\ref{fig:diffusion_temperature}(b), (d) and (f), we show the temperature distribution at $x = 0$ as a function of $y$: one sees that temperature decays slower in systems with a larger $\kappa$ or $l_\text{diff}$.

\section{Summary  and discussion}
\label{sec:summary}
To summarize, we have investigated  viscous hydrodynamic transport of electrons in a clean two-dimensional sample between two semi-infinite planes separated by an impenetrable barrier with a finite-size slit. We have calculated the velocity profile assuming that a driving voltage is applied between these planes.  We demonstrate that the standard boundary conditions at the barrier, say ``no-stress'', together with the condition on the given voltage drop across the sample cannot uniquely determine the velocity. There exist infinitely many solutions satisfying these conditions. Among all possible solutions, we have found the unique one that leads to a finite total dissipated power, thus respecting the energy conservation.   
Our central result is that this only physical solution, which is characterized by a finite  current through the slit and yields finite total dissipation, obeys \textit{simultaneously} both the ``no-slip'' and ``no-stress'' boundary conditions. This in turn means that the slip length, which is often used for the partial-slip (Maxwell) boundary condition, is irrelevant for this problem and does not  enter any physical result.

Several comments are in order here to compare this result with previous theoretical studies.  The  resistance of this  ``no-slip, no-stress'' solution  equals to  the ``superballistic'' resistance obtained in Ref.\,\cite{GuoPNAS17} for the same system with the no-slip boundary condition. The distribution  of the velocity calculated in Ref.~\cite{GuoPNAS17} coincides with the corresponding distribution in our  solution. 
However, the fact that this solution also satisfies the no-stress condition \eqref{eq:no-stress} was not mentioned there. At the same time, in Ref.\,\cite{PershogubaYongGlazmanPRB20}, a solution was found, which satisfies the no-stress condition \eqref{eq:no-stress}, but does not satisfy the no-slip condition  \eqref{eq:no-slip}. As we demonstrated above, this solution yields a divergent total dissipation and, therefore, is nonphysical.

Above, we developed theory for an infinitely thin barrier.  It would be very instructive to get solution for a wall with a finite thickness $a$ and then take the limit $a\to 0$. Unfortunately, the obtaining of such a solution is very tricky and is out of the scope of the current paper.  However, we expect that the solution for a finite but very small $a$ ($a \ll w$) would be very close to our solution for  $a=0$.
The point is that our solution not only yields finite dissipation but also allows one to calculate resistance of the sample in three different ways: via evaluating the total viscosity-induced dissipation power; (ii) via the definition of the resistance $V/I$, (iii) via evaluating the total work done by the electric force.  All three ways yield the same value for resistance, so our solution is fully consistent. Any solution which is regular for finite $a$ but becomes singular in the limit $a\to 0$ should give a very large 
 resistance, calculated  from viscosity-induced dissipation, inconsistent with the   
 value $V/I.$

We have used the obtained velocity flow to solve the heat balance equation 
and investigate evolution of the electronic temperature profile while changing the heat conductivity. We have found that the overheated areas appear near the slit endpoints. The shape of these areas is determined by a competition between heat diffusion and heat convection. For a large heat conductivity, the diffusion dominates and these areas are approximately symmetric with respect to the barrier separating the two half-planes. In addition,
by increasing the heat conductivity, we see a crossover between the ring-shape and the extended temperature profiles.
Increasing the voltage (and, hence, the role of convection) leads to two effects: enhancement of asymmetry of the temperature pattern along the lines of the hydrodynamic flow and increase of the overheating. In the limit of zero heat conductivity, the temperature at the  slit endpoints diverges. This divergence is cured by a finite heat conductivity.  Our predictions can be directly verified experimentally with the use of recently developed ultra-precise  techniques for simultaneous nanoimaging of temperature and current distributions 
\cite{FinklerNanoLett10,Vasyukov,HalbertalNat16,HalbertalScience17,Zeldov2019,VoolX2020,UriNatPhys20,AnahoryNanoscale20,AmitNature21,AmitZeldovNature22}.

\begin{acknowledgments}
We thank Igor Gornyi as one of the authors of the original idea, for an enormous contribution to this paper and his unmeasurable kindness. We also thank Igor Burmistrov for the idea to use Keldysh-Sedov theorem in the analysis of this problem.
The work of VK  was supported
by the Russian Science Foundation (Grant No. 20-12-00147).

\end{acknowledgments}

\appendix

\section{Potential, vorticity, and velocity from Keldysh-Sedov's formula}
\label{ks_app}

In this appendix, we derive the family of solutions for the velocity profile that satisfies the no-stress boundary condition on the walls.
Following Ref.\,\cite{PershogubaYongGlazmanPRB20}, we take as the boundary conditions (i) Eq.~\eqref{eq:epotential_bc} for electric potential at infinity; (ii) the symmetry requirement for the electric potential in the slit $\phi = V/2$, and (iii) the no-stress boundary condition, which results in the condition of zero vorticity on the walls: 
\begin{equation}
\omega=\left(\frac{\partial v_y}{\partial x} -  \frac{\partial v_x}{\partial y}\right)\Big|_{|x|>w,y \rightarrow 0} = 0
\end{equation}
With these conditions, we get for the complex function in Eq.~(\ref{fz-def})
\begin{equation}
    \begin{split}
       \text{Re}f(z)&|_{|x|<w, y=0} = -\frac{enV}{2\eta},  \\
       \text{Im}f(z)&|_{|x|>w,y \rightarrow 0} = 0, \\
       & f(z)|_{r\rightarrow \infty} = 0, 
    \end{split}
    \label{eq:boundary_1}
\end{equation}
and for the velocity field
\begin{equation}
\begin{split}
& v_y = 0|_{|x|>w}, \\
& v_x = 0|_{|x|<w}.
\end{split}
\label{eq:boundary2}
\end{equation}
Thus, the problem is reduced to finding a holomorphic function in the complex half-plane with the known real part on the segment $y=0$, $-w<x<w$ and imaginary part on the $y=0$, $|x|>w$. The solution to this problem is given by the Keldysh-Sedov theorem \cite{TFKP}. 

Let us state this theorem here in its complete form, following Ref.~\cite{TFKP}. To begin, we formulate the general problem addressed in the theorem. Consider $2n$ points on the real axis such that $-\infty<a_1<b_1<..<a_n<b_n<\infty$ and two real functions $u(x),\;v(x)$ with a finite number of the first-order discontinuity points. The function $u(x)$ is defined on all intervals $(a_k, b_k)$ and $v(x)$ -- on all intervals $(b_k, a_{k+1})$ (with $k=1,2,...,n$ and $a_{n+1}\equiv a_n$). The problem is to find a function $f(z)$, analytic in the upper half plane such that $\mathrm{Re}f(z)=u(x)$ on the intervals $(a_k, b_k)$ and $\mathrm{Im}f(z)=v(x)$ on the intervals $(b_k, a_{k+1})$. 

\textbf{Theorem}. The mixed problem for the upper half-plane has a unique solution $f(z)$, satisfying the following conditions:
(i) $f(z)$ is limited in the vicinity of all $a_k$; (ii) the integral $\int^z f(z) dz$ is limited in the vicinity of all $b_k$, and (iii) the limit $f(\infty)$ is finite and assumed to be real for simplicity. The unique solution is given by the following formula:
\begin{equation}
    f(z)=\frac{1}{\pi i g(z)}\sum_{k=1}^{n}\left(U_k(z)+iV_k(z)\right)+ \frac{f(\infty)}{g(z)},
\end{equation}
where 
$$
g(z)=\prod_{k=1}^{n}\frac{z-b_k}{z-a_k},
$$
and
$$
U_k(z)=\int_{a_k}^{b_k}\frac{u(t)g(t)dt}{t-z},\quad V_k(z)=\int_{b_k}^{a_{k+1}}\frac{v(t)g(t)dt}{t-z}.
$$

Applying the Keldysh-Sedov theorem to our case, we conclude that there are no solutions for $f(z)$ that are finite near points $(-w, 0)$ and $(w, 0)$. Furthermore, all possible solutions $f(z)$ analytic in the upper half-plane under conditions of Eq. (\ref{eq:boundary_1})
with finite $\int\limits_{}^{z} f(z) \, dz $ read as follows:
\begin{equation}
    f(z) = \frac{NeV}{2\eta} \left[ -1 + \frac{z}{\sqrt{z^2 - w^2}}\right] + \frac{\mathcal{C}_1}{\sqrt{z^2-w^2}}.
    \label{eq:fz1}
\end{equation}

We may now compute the velocity profile. Using the continuity equation $\mathbf{\nabla} \cdot \mathbf{v} = 0$ for incompressible liquid and definition of vorticity, one arrives at the expression
\begin{equation}
   \partial_{\bar{z}}( v_y + i v_x ) = \frac{1}{2} \text{Im} f(z).
\end{equation}
Then, by definition,
\begin{equation}
 u(z, \bar{z}) \equiv v_y + i v_x = \frac{1}{2}\int\limits_{}^{\bar{z}} [\text{Im} f](z. \bar{z}) \, d\bar{z} +C(z),
\end{equation}
where $C(z)$ is an arbitrary function, holomorphic in the upper half-plane. 
We can see that the part proportional to $\mathcal{C}_1$ leads to the logarithmic growth of velocity at infinity. Therefore we should put $\mathcal{C}_1=0$. 
Equation (\ref{eq:boundary2}) fixes the imaginary part of $C(z)$ on the segment $y=0$, $-w<x<w$ and real part on the segment $y=0$, $|x|>w$.  This allows us to once again use Keldysh-Sedov's formula to compute $C(z)$ and arrive at Eq.~\eqref{eq:umain}
with $\mathcal{C}$ an undetermined constant. Recall that in Eq.~\eqref{eq:umain}, we have chosen the branch cuts to run along the screen. 

Let us note that real and imaginary parts of $f(z)$ deliver non-unique solutions to the Laplace equations $\Delta \phi = 0$ and $\Delta \omega = 0$, correspondingly. This non-uniqueness is related to the fact that the solutions diverge near the edges of the slit.

\section{  Alternative method of calculation: regular and singular solutions }
\label{even_more_singular_app}

In this appendix, we briefly describe a  method allowing us to obtain general solutions
obeying    no-slip and
no-stress boundary conditions     (a more detailed discussion will be published elsewhere). The method  is based on    conformal  transformation,
\be
z= w  \cosh \left( \frac{\pi  z_1}{w}\right),
\label{map}
\ee
which maps complex plane    $z_1=x_1+ i y_1$ into the  complex  plane    $z=x+ i y$ of our problem.
Slit geometry in coordinates  $(x,y)$   corresponds to strip geometry in $(x_1,y_1)$
coordinates:   $ 0< y_1< w,$ $ -\infty <x_1<\infty.$    The Stokes equation, Eq.~\eqref{eq:stokes_eq},
describing flow of the incompressible fluid with 
$ (\mathbf{\nabla} \cdot \mathbf v)=0,$
dramatically simplifies in  the new coordinates:
\be
\frac{\partial \mathcal{F}}{\partial \bar{z}_1}=0,
\label{dWdz}
\ee
where complex potential is introduced,
\be
\mathcal{F}= \frac{\pi^2 }{J}\omega_1 + i \frac{eN}{\eta}\phi,
\label{W}
\ee
with $\phi$ for the electrical potential, and
$\omega_1$ for the vorticity  of the field  $\mathbf u = \pi^{-1} \sqrt{J} \mathbf v_1$. Here, $\mathbf v_1$ is
the velocity field in new coordinates and 
$$J=\det(\p x_\alpha/\p x_{1\beta})=
\pi^2 \sinh (\pi z_1/w) \sinh (\pi \bar{z}_1/w)$$ 
is the Jacobian of the map.
Any function $\mathcal{F}=\mathcal{F}(z_1)$ of a complex variable $z_1$ solves  Eq.~\eqref{dWdz}.
The imaginary part
of this function yields the electrical potential, while the real part gives the vorticity up to the known factor $J/\pi^2$.

To find velocity profiles starting from a given function $\mathcal{F}(z_1)$, one  has to find stream
function $\psi$ of the field $\mathbf u$  by solving  equation  $\omega_1=\Delta_1 \psi.$ The solution
depends on a certain function $A(z_1)$ that should be found from the boundary conditions.

Let us give several examples of solutions corresponding to  the fixed potential drop
between $y=-\infty$ and $y =\infty,$ and obeying both no-slip and no-stress boundary conditions.
 For simplicity, we use below the units where all lengths are measured in units of  $w,$ complex potential is measured in units of total voltage drop $V,$ and velocity is measured in units  $eNV w/4\eta. $
The regular solution, discussed in the main body of the text corresponds to
\be
 \mathcal{F}^{(1)}(z_1) =-(i/2) \coth (\pi z_1/w).
\label{W1S1}
\ee
   This solution
gives a regular distribution of the velocity in the slit:
\be
v_y^{(1)} (x)= \sqrt{1-x^2}.
\label{vy-for-y=0}
\ee
The distribution of the velocity in the whole system corresponding  to a no-slip$+$no-stress solution with complex potential $\mathcal{F}^{(1)}$  in the real form reads
\be 
\begin{aligned}
&v^{(1)}_x=\frac{ r(1-r^2)\sqrt{\mathcal R^2-1}~ \mathrm{sign}(y) }{ \mathcal R^2-r^2},
\\
&v^{(1)}_y=\frac{ \mathcal R (1-r^2)^{3/2}}{ \mathcal R^2-r^2},
\end{aligned}
\label{v1xy}
\ee
where
\be
\begin{aligned}
&\mathcal R = (\rho_+ + \rho_-)/2, \quad r= (\rho_+ - \rho_-)/2,
\\
&\rho_+=\sqrt{(1+x)^2+y^2}, \quad \rho_-=\sqrt{(1-x)^2+y^2}.  
\end{aligned}
\ee
One can  easily check 
that Eq.~\eqref{v1xy} yields Eq.~\eqref{vy-for-y=0}

The same fixed voltage
drop can also be reproduced by complex potential 
$$  \mathcal{F}^{(n)}(z_1) =-(i/2)  \coth^n (\pi z_1/w) $$ with any positive odd integer $n$.
It turns out that solutions with $n\geq 3$ are characterized by non-integrable vorticity and thus cannot be captured by the Keldysh-Sedov formula from Appendix~\ref{ks_app}.
For example, for $n=3$ we have
$$v_y^{(3)}=   \sqrt{1-x^2} + \frac{1}{\sqrt{1-x^2}}.$$  
Corresponding velocity  distribution becomes
\be 
\begin{aligned}
&v^{(3)}_x=\frac{ r(1-r^2)\sqrt{\mathcal R^2-1}(3 \mathcal R^4+r^4-3 \mathcal R^2-r^2)~ \mathrm{sign}(y) }{(\mathcal R^2-r^2)^3},
\\
&v^{(3)}_y=\frac{ \mathcal R (1-r^2)^{3/2}(3 \mathcal R^4+r^4- \mathcal R^2-3r^2)}{ (\mathcal R^2-r^2)^3}.
\end{aligned}
\label{v3xy}
\ee
We see that
$v_y^{(3)}$
has integrable singularity
at the slit edges.  Hence, both $n=1$ and $n=3$ cases correspond to a finite current through the slit. 
Both these solutions are captured by the method developed in the main text and can be obtained from Eqs.~(\ref{eq: general boundary v}) and (\ref{eq: convolution}). At the same time, 
the solution with $n=3$ yields non-integrable vorticity and infinite dissipation and  should be discarded. 

The solution corresponding to  ${n=5}$ has a nonintegrable velocity distribution inside the slit. In this case, we have
$$v_y^{(5)}=  \sqrt{1-x^2} + \frac{2}{\sqrt{1-x^2}} - \frac{1}{3 (1-x^2)^{3/2}}.$$
This solution has strong singularity at the slit edge and, therefore, cannot be obtained within the integral-equation method used in the main body of the paper.
Needless to say, the total dissipated power for this solution is divergent.

Regular solution $\mathbf v^{(1)}$ and singular solutions $\mathbf v^{(n)}$ ($n=3,5,\dots$) simultaneously obey no-slip and no-stress boundary conditions. This implies certain choice of  function $A^{(n)}(z_1)$ for solution with a given value of $n.$  One  can choose  function   $A^{(1)}(z_1),$  corresponding to complex potential $\mathcal F^{(1)}$ in a different way, so the solution for velocity would obey the no-stress boundary condition only. Specifically, such a solution reads
\be 
\begin{aligned}
&\delta v_x=\frac{ r \sqrt{\mathcal R^2-1}~ \mathrm{sign}(y) }{\mathcal R^2-r^2},
\\
&\delta v_y=\frac{ \mathcal R \sqrt{1-r^2}}{ \mathcal R^2-r^2}.
\end{aligned}
\label{dvxy}
\ee
Exactly at the barrier, the $x-$component of velocity is non-zero
\be
\delta v_x (x,y \to 0)= \frac{\mathrm{sign}(x)\mathrm{sign}(y)}{\sqrt{x^2-1}}. 
\ee
For $y=0, ~|x|<1 $, we find for the $y$ component of velocity
\be
\delta v_y (x,0)= \frac{1}{\sqrt{1-x^2}}. 
\label{dvy}
\ee 
It is worth noticing that solution $\delta \mathbf v$ corresponds to zero vorticity and electrical potential.  
Let us now consider a general   no-stress solution which has singularity not higher than $(1-x^2)^{-1/2}$: 
\be
\mathbf v^{\rm no-stress}=
a^{(1)}  \mathbf v^{(1)}+
a^{(3)}  \mathbf v^{(3)}+
\xi   \delta \mathbf v.
\ee
Comparing with   Eq.~\eqref{general_noslip}, we  find that 
\be
\mathbf v^{(1)}=\mathbf u_1, ~
\mathbf v^{(3)}=\mathbf u_1+\mathbf u_2,~
\delta \mathbf v= \mathbf u_2+\mathbf u_3,
\label{correspondence}
\ee
so
\be
c_1=a^{(1)} +a^{(3)},\quad c_2= a^{(3)}+ \xi, \quad  c_3= \xi.
\ee
For the solution given by   Eq. \eqref{eq:umain} we have \be a^{(1)}=1, \quad a^{(3)}=0, \quad \xi= \frac{\mathcal C-1}{2},
\label{aaa}
\ee so
\be
c_1= 1, \quad c_2=c_3= \frac{\mathcal C-1}{2},
\label{ccc}
\ee
and we reproduce  Eq.~\eqref{ci-for-Glazman}.  
Using Eqs.~\eqref{v1xy},   \eqref{v3xy}, \eqref{dvxy}, and Eq.~\eqref{S}, we can  analyze surface contribution to total dissipation coming from the  second term in Eq.~\eqref{P1+P2}. For solution $\mathbf v^{(1)}$, the corresponding contribution to $\mathcal P$ equals to zero. Contributions coming from both $\mathbf v^{(3)}$ and $ \delta \mathbf v$  are singular. In particular,  for the solution given by   Eq. \eqref{eq:umain}, where $a^{(3)}=0$ 
 [see  Eq.~\eqref{aaa}], we find  for $\rho \to 0$: 
$$ \mathbf S \propto \xi^2 \frac{\boldsymbol{\rho}}{\rho^3 }.$$ 
Then, we get for flux  through the contour close to the slit end: $\mathbf S ~ 2\pi \boldsymbol{\rho} \propto \xi^2/\rho. $ This expression diverges for $\xi \neq 0.$

\begin{widetext}
\begin{center}
    \begin{table}[h!]
    \centering
    \begin{tabular}[b]{|*{6}{c|}}
\hline 
& $v_y(|x|<1,0)$ & $v_x(|x|>1,y \to \pm 0)$ & $\mathcal{F}$ - complex potential & Voltage drop & $\omega$ - \text{vorticity} \\ \hline
$\mathbf {v}^{(1)}$ & ${\sqrt{1-x^2}}$ & $0$ & $\mathcal{F}_1 = -(i/2)\coth{\pi z_1} $ & 1 & $\omega_1 \propto \rho^{-\frac{1}{2}} $ \\ \hline
$\mathbf {v}^{(3)}$ & $ \sqrt{1-x^2}+\frac{1}{\sqrt{1-x^2}}$ & $0$ & $\mathcal{F}_3 = -(i/2)\coth^3{\pi z_1} $ & 1 & $\omega_3 \propto \rho^{-\frac{3}{2}} $ \\ \hline
$\mathbf {\delta v}$  & $\frac{1}{\sqrt{1-x^2}}$ & $\frac{\text{sign}(x)\text{sign}(y)}{ \sqrt{1-x^2}}$ & 0 & 0 & 0  \\ \hline
$\mathbf{u}_1$ & \scalebox{0.7}{$\sqrt{1-x^2}$} & $0$ & $\mathcal{F}_1 $ & 1 & $\omega_1\propto \rho^{-1/2}$\\ \hline
$\mathbf{u}_2$ & $\frac{1}{\sqrt{1-x^2}}$ & $0$ & $\mathcal{F}_3 - \mathcal{F}_1$ & 0 & $\omega_3 - \omega_1 \propto \rho^{-\frac{3}{2}}$ \\ \hline 
$\mathbf {u}_3$ & 0 & $\frac{\text{sign}(x)\text{sign}(y)}{\sqrt{1-x^2}}$ & $\mathcal{F}_1-\mathcal{F}_3$ & 0 & $\omega_1 - \omega_3 \propto - \rho^{-\frac{3}{2}}$ \\ \hline 
\end{tabular}
\caption{Specific properties of different solutions.}
    \label{table}
\end{table}
\end{center}
\end{widetext}

Before closing this appendix, let us make several  comments on interesting properties of velocity flows described above.   First of all, one can check that solution $\delta\mathbf v$ has  exactly zero vorticity and corresponds to zero electrical potential in the whole space. We also  note that solution $\mathbf u_3$ yields zero flux through the slit, $u_{3y} (|x|<1,0) \equiv 0 .$  Solutions $\mathbf u_2$ and $\mathbf u_3$   are produced  by potentials $\delta \mathcal F$ and  $-\delta \mathcal F,$ respectively, where $\delta \mathcal F=\mathcal F^{(3)}- \mathcal F^{(1)}$ is evidently different from the potential $\mathcal F^{(1)}$  that produces regular solution $\mathbf u_1.$ At the same time, voltage drops corresponding to   $\mathbf u_2$ and $\mathbf u_3$ are exactly equal to zero, because $(\mathcal F^{(3)}- \mathcal F^{(1)})|_{x\to \pm \infty}=0.$   Specific properties of different solutions are summarized in Table~\ref{table}.

\section{Solution with the partial-slip boundary condition: Finite slit vs. point source}
\label{app:the_missing_solution}

In the main text, we have obtained a solution that satisfies both the no-slip (\ref{eq:no-slip}) and no-stress (\ref{eq:no-stress}) boundary conditions.
This solution however
  does not coincide with solution in  Ref.~\cite{PershogubaYongGlazmanPRB20},
where the point-source solution 
from Ref.~\cite{FalkovichLevitovPRL17} was invoked for deriving the velocity profile in the finite-slit geometry with the no-stress boundary condition. In this appendix, we explicitly show why the solution for $w \equiv 0$
is inadequate for finding the velocity profile for a finite slit $w>0$.

We will follow the derivation of Ref.~\cite{FalkovichLevitovPRL17} and require the velocity profile to satisfy (i) the ``partial-slip'' boundary condition Eq.~\eqref{eq:boundary_condition} at barriers (i.e., $y = 0$ for $|x| > w$), and (ii)  zero $v_x$ in the slit (i.e., $y = 0$ for $|x| < w$).
Rewriting these conditions through the stream function $\psi(x,y)|_{y=0}$, 
we have 
\begin{equation}
    \begin{cases}
         \dfrac{\partial \psi}{\partial y} = \lambda \dfrac{\partial^2 \psi}{\partial y^2}, \quad &(y = 0,\ \  |x| \geq w), \\[0.2cm]
         \dfrac{\partial \psi}{\partial y} = 0, \quad &(y = 0, \ \ |x| < w).
    \end{cases}\,
    \label{eq:boundaries for psi}
\end{equation}
We can cast Eq.~\eqref{eq:boundaries for psi} in terms of the Fourier transformed stream function $\tilde{\psi}(k, y)$, 
leading to
\begin{align}
\frac{\partial \tilde{\psi}(k, y)}{\partial y}& = \lambda \frac{\partial^2 \tilde{\psi}(k, y)}{\partial y^2}
\notag
\\
&-\frac{\lambda}{\pi} \int dk_1 \frac{\sin[(k-k_1)w]}{k-k_1} \frac{\partial^2 \tilde{\psi}(k_1, y)}{\partial y^2}\Big{|}_{y=0}.
\label{eq:mixed_bc_another_form}
\end{align}
Substituting the general form of the solution, Eq.~\eqref{eq: fourier_of_velocities}, into Eq.~\eqref{eq:mixed_bc_another_form} we arrive at
\begin{equation}
\begin{aligned}
&  |k| \mathcal{A}_1(k) - \mathcal{A}_2 (k)   + \lambda [ k^2 \mathcal{A}_1(k) - 2 |k| \mathcal{A}_2(k)]
\\
&=  \frac{\lambda}{\pi} \int \frac{dk_1}{ k - k_1 } \sin[(k-k_1) w] [ k_1^2 \mathcal{A}_1 (k_1) - 2 |k_1| \mathcal{A}_2(k_1) ].
\end{aligned}
\label{eq:mixed_bc_continue}
\end{equation}
By setting $w \equiv 0$ in Eq.~\eqref{eq:mixed_bc_continue}, we reproduce the relation from Ref.~\cite{FalkovichLevitovPRL17} (see Sec.~II of its Supplementary Material):
\begin{equation}
    \mathcal{A}_2(k) = \frac{|k| + \lambda k^2}{1 + 2 |k| \lambda} \mathcal{A}_1(k).
    \label{eq:c1c2lambda}
\end{equation}
Note that this relation is different from
\begin{equation}
    \mathcal{A}_2(k) = |k| \mathcal{A}_1(k),
\label{ourc1c2}
\end{equation}
which corresponds to $v_x(x,0) = 0$ at the wall (although they become equivalent for $\lambda=0$).
 This, in particular, means that the asymptotic velocity profile at $r\to \infty$ is different for the two solutions, if $\lambda\neq 0$. For example, in the limit $\lambda\to \infty$, referred to as the no-stress case in Ref.~\cite{FalkovichLevitovPRL17}, one obtains $\mathcal{A}_2(k)=|k|\mathcal{A}_1(k)/2$ from Eq.~\eqref{eq:c1c2lambda}.
This difference in the factor of two is exactly the one that determines the difference between $\mathcal{C}=w^2$ in our solution and $\mathcal{C}=2 w^2$ in Ref.~\cite{PershogubaYongGlazmanPRB20}. Indeed, Ref.~\cite{PershogubaYongGlazmanPRB20} used the large-$r$ asymptotics of the ``point-source'' solution (\ref{eq:c1c2lambda}) to fix the constant $\mathcal{C}$ for the no-stress boundary conditions. We  also notice that Eq.~\eqref{eq:c1c2lambda} has non-integrable divergence of velocity $v_x\propto 1/x$ at $x\to 0$ and generates divergent dissipation.

On a technical level, the key point is that
the kernel in the last (integral) term in Eq.~\eqref{eq:mixed_bc_continue}
acts as an effective delta-function for
any continuous velocity profile $v_y(x,y=0)$ in our geometry.
For such functions, it holds that 
\begin{equation}
\frac{1}{\pi}\int \frac{dk_1}{k - k_1 } \sin[(k-k_1)w] k_1^2 \mathcal{A}_1 (k_1) = k^2 \mathcal{A}_1 (k),
     \label{eq:c1_integral}
\end{equation}
and similarly for the term with $\mathcal{A}_2(k)$
in Eq.~\eqref{eq:mixed_bc_continue}.
With Eq.~\eqref{eq:c1_integral}, the $\lambda$-dependent terms on both sides of Eq.~\eqref{eq:mixed_bc_continue} cancel. We then immediately see that our solution (\ref{ourc1c2}) with $v_x(x,0) = 0$, which was originally obtained for the no-stress condition (\ref{eq:no-stress}), is a valid solution for arbitrary $\lambda$. This solution is, however, missed, when one sets $w\equiv 0$ before performing the integration over $k_1$, thus completely disregarding the integral term in Eq.~\eqref{eq:mixed_bc_continue}.  Representing the no-stress solution for a finite-slit problem ($w>0$) through the point-source solution~\eqref{eq:c1c2lambda} from Ref.~\cite{FalkovichLevitovPRL17} with $\lambda\to \infty$, as was done in Ref.~\cite{PershogubaYongGlazmanPRB20}, is thus not justified. 

One can easily demonstrate that  Eq.~\eqref{eq:c1_integral} holds for our regular solution for which velocity in the slit is  given by Eq.~\eqref{eq: general boundary v} with $c_1=1, c_2=0.$
Using general equations derived in Appendix \ref{app:derivation of the Kernels}, one can find that the Fourier transform   $\tilde{\mathcal{A}}_1(x)$ of $\mathcal{A}_1(k)$ for the regular solution reads: $\tilde{\mathcal{A}}_1(x) =0 $ for $x<-w,$  $\tilde{\mathcal{A}}_1(x)=\int_{-w}^x  dx' v_y(x',0)$ for $|x|<w$ and $\tilde {\mathcal{A}}_1(x)= I w/eN,$ for $x>w.$ Here, $I$ is the total current through the slit. By taking the inverse Fourier transform of this expression, and using the identity 
$\int_{-\infty}^\infty dk_1  e^{ik_1 x}{\sin[(k-k_1)w]}/({k-k_1})   
=  ({\pi}/{2}) e^{ikx} [\text{sgn}(w-x) + \text{sgn}(w+x)]$,
one can  get the following  analytical expression for   
$\mathcal{A}_1 (k)$ (see the Supplementary Information~\cite{Supmat} for details):
\begin{equation}
    \mathcal{A}_1 (k) =-i \frac{2I}{e N} \frac{J_1( k w )}{k^2},
\label{eq:c1k_explicit}
\end{equation}
where $J_1(k w)$  is  the Bessel function of the first kind. Substituting Eq.~\eqref{eq:c1k_explicit} into 
Eq.~\eqref{eq:c1_integral} one can verify  Eq.~\eqref{eq:c1_integral}   by straightforward  calculation.  In the Supplementary Information,  we also prove Eq.~\eqref{eq:c1_integral}    
  for any continuous velocity profile $v_y(x,y=0)$ in our geometry without appealing to  specific expressions for $v_y(x,0)$ and $\mathcal{A}_1 (k)$.

\section{Work of the electric field}
\label{app:extra_electric_work}

In the main text we have shown that with our solution
the total viscosity-induced dissipation is equal to the work of the electric field. The corresponding resistance equals to that obtained by the resistance definition~\cite{GuoPNAS17} for the no-slip boundary conditions. In this appendix, we consider the work of the electric field obtained with the choice $\mathcal{C} = 2w^2$~\cite{PershogubaYongGlazmanPRB20} for the velocity profile (\ref{eq:umain}).

As the starter, the electric field of our case agrees with that of Ref.~\cite{PershogubaYongGlazmanPRB20}: only the velocity profiles are different.
As we have already evaluated the work of the electric field of our case, in this appendix we only need to figure out the extra work induced by the velocity difference between our solution and that of Ref.~\cite{PershogubaYongGlazmanPRB20}.
The local extra work equals
\begin{equation}
    N e \delta \mathbf{v} \cdot \nabla \phi = -i\left( \delta \bar{u} \frac{\partial}{\partial z} - \delta u \frac{\partial}{\partial \bar{z}} \right) ne \phi,
\end{equation}
where $\delta \mathbf{v}$ refers to the difference in velocity between the solution of Ref.\,\cite{PershogubaYongGlazmanPRB20} and that of the current paper, and $\delta u = \delta v_y + i \delta v_x$.

The total difference of the work done by the electric field in the bulk of the sample is given by
\begin{equation}
\begin{aligned}
&- i \iint dx dy \left( \delta \bar{u} \frac{\partial}{\partial z} - \delta u \frac{\partial}{\partial \bar{z}} \right) ne \phi \\
&= \iint d\theta dr r\frac{e^2n^2V^2 }{16 \eta } \frac{w^2(w^2 - r^2 \cos 2\theta)}{[r^4 + w^4 - 2 w^2 r^2 \cos(2\theta)]^{3/2}}\\
&=   \frac{\pi}{16} \frac{(e N V w)^2}{\eta},
\end{aligned}    
\end{equation}
which is twice as that of Eq.~\eqref{eq:superballistics_dissipation} of the superballistic case.
Using this result, with the velocity profile of Ref.~\cite{PershogubaYongGlazmanPRB20}, the obtained resistance equals to
\begin{equation}
    G_{2w^2} = \frac{3}{2} G_{\text{no-slip}},
\end{equation}
if evaluated by the total work done by the electric field in the bulk of the sample,
where $G_\text{no-slip} = 1/R_\text{no-slip}$ is the superballistic conductance calculated in Ref.~\cite{GuoPNAS17} for no-slip boundary conditions [see Eq.~\eqref{eq:viscous_resistance} of Sec.~\ref{sec:velocity}]. This result, however, disagrees with the conductance obtained in Ref.\,\cite{PershogubaYongGlazmanPRB20} via the definition of the resistance $R = V/I$, where $G_{2w^2} = 2 G_{\text{no-slip}}$ instead.
Thus, for the velocity profile (\ref{eq:umain}) with $C=2w^2$, the resistances obtained by the three different methods ($V^2/\mathcal{P}$, $V/I$, and $V^2/W$) introduced in Sec.~\ref{sec:velocity} are all different. 
This should be contrasted with the results of Sec.~\ref{sec:velocity} for the finite-dissipation profile with $C=w^2$, for which all these three methods consistently yield the same resistance.

\section{Derivation of the kernels for the integral equations}
\label{app:derivation of the Kernels}

To obtain the kernels in Eq.~(\ref{eq:Kernels}), we apply a standard two-dimensional hydrodynamic technique and introduce the stream function $\psi(x,y)$, such that 
\begin{align}
    v_x = \frac{\partial \psi}{\partial y},
    \quad
    v_y= -\frac{\partial \psi}{\partial x}.
\end{align}
The Laplace equation $\Delta \omega = 0$ is then equivalent to the biharmonic equation $$\Delta^2 \psi = 0.$$  
The Fourier transform with $$\tilde{\psi}(k,y)=\int\psi(x,y) e^{-ikx} dx$$ 
leads us to the equation
\begin{equation}
\frac{\partial ^4 \tilde{\psi}(k, y)}{\partial y^4}-2k^2\frac{\partial^2 \tilde{\psi}(k, y)}{\partial y^2} + k^4\tilde{\psi}(k, y)=0.
\label{eq:fourier psi}
\end{equation}
Since we are looking for the solution in the upper half-plane, we should keep only the partial solutions for which $\tilde{\psi}(k, y) \rightarrow 0$ when $y \rightarrow \infty$. 
Hence
\begin{equation}
\begin{aligned}
& \tilde{\psi}(k,y) = \mathcal{A}_1 e^{-|k|y} +  \mathcal{A}_2  y e^{-|k|y}, \\
& \tilde{v}_x(k, y) = -|k|\mathcal{A}_1\, e^{-|k|y}+(1-|k|y)\mathcal{A}_2\, e^{-|k|y}, \\
& \tilde{v}_y(k, y) = -ik\left(\mathcal{A}_1 e^{-|k|y} +  \mathcal{A}_2  y e^{-|k|y}\right).
\end{aligned}
\label{eq: fourier_of_velocities}
\end{equation}
Here $ \tilde{v}_x, \tilde{v}_y$ denote the Fourier transforms of velocities 
\begin{align}
    \tilde{v}_x(k, y) =\int v_x(x,y) e^{-ikx} dx, 
    \notag
    \\
    \tilde{v}_y(k, y)=\int v_y(x,y) e^{-ikx} dx.
    \notag
\end{align}
It is now easy to express $\mathcal{A}_1$ and $\mathcal{A}_2$ through $ \tilde{v}_x(k, 0)$ and $\tilde{v}_y(k, 0)$:
\begin{equation}
\begin{aligned}
\mathcal{A}_1 & = i \frac{\tilde{v}_y(k, 0)}{k} \\
\mathcal{A}_2 & = \tilde{v}_x(k, 0)+\tilde{v}_y(k, 0) \frac{i|k|}{k}
\end{aligned}
\label{eq:c_1 and c_2}
\end{equation}
Expressing then $\tilde{v}_x(k, y)$ and $\tilde{v}_y(k, y)$ through 
$ \tilde{v}_x(k, 0)$ and $\tilde{v}_y(k, 0)$, and performing the inverse Fourier transformation, we obtain
\begin{equation}
\begin{aligned}
&\vec{K}_1(x-x', y)=\int \frac{dk}{2\pi}
e^{-|k|y+ik(x-x')}  
\begin{pmatrix}
&1-|k|y\\[0.2cm]
&-iky 
\end{pmatrix},  \\[0.5cm]
& \vec{K}_2(x-x', y)= 
\int \frac{dk}{2\pi}
e^{-|k|y+ik(x-x')}  \begin{pmatrix}
&-iky &\\[0.2cm]
&1+|k|y
\end{pmatrix},
\end{aligned}
\end{equation}
which yields Eq.~(\ref{eq:Kernels}) of the main text.

\section{Finding velocity on the walls through Chebyshev polynomials}
\label{strict proof for velocity}

In this appendix, we demonstrate the uniqueness of the non-divergent solution 
$$v_x(|x|>1, y=0) = 0$$
for the arbitrary $\lambda$ boundary condition (\ref{eq:boundary_condition}) with a finite $\lambda$. This is done by using an expansion of the velocity in Chebyshev polynomials (cf. Ref.~\cite{PershogubaYongGlazmanPRB20}).
It is possible to rewrite the equation for $v_x$ using the new variable $\alpha = w/x$ in the following form:
\begin{equation}
v_x(\alpha,0) = \frac{4 \lambda \alpha^2}{\pi w} \frac{\partial}{\partial \alpha}  \fint_{-1}^{1} \frac{1}{\alpha'-\alpha} v_x(\alpha',0)d\alpha'.
\label{eq:better inteq for v_x}
\end{equation}
If $v_x(\alpha)/\sqrt{1-\alpha^2}$ is integrable inside the slit, we can expand the velocity $v_x(\alpha, 0)$ in the series:
\begin{equation}
    v_x(\alpha, 0)=\sqrt{1-\alpha^2}\, \sum_{n=0}^{\infty} C_n U_{2n+1}(\alpha).
\label{eq:new form of v_x}
\end{equation}
Here $U_n(x)$ is the Chebyshev polynomial of the second kind, which satisfies the following relations:
\begin{equation}
\begin{aligned}
 \fint_{-1}^{1} \frac{\sqrt{1-x^2}U_n(x)}{x'-x} & dx'  = -\pi T_{n+1}(x),\\
 \frac{\partial T_n(x)}{\partial x} &= n U_{n-1}(x),
\end{aligned}    
\end{equation}
where $T_n(x)$ is the Chebyshev polynomial of the first kind.
From the first line of \eqref{eq:new form of v_x} we get
\begin{equation}
    \sum_{n=0}^{\infty} \left[\sqrt{1-\alpha^2}+\frac{8 \lambda \alpha^2}{w}(n+1)\right] C_n U_{2n+1}(\alpha) = 0.
\end{equation}
Using the definition of Chebyshev polynomials through trigonometric functions,
\begin{equation}
    U_n(\alpha) = \frac{\text{sin}[(n+1)\theta]}{\text{sin}(\theta)},\quad \alpha = \text{cos}(\theta),\quad \theta \in [0, \pi],
    \label{eq: Chebyshev polynomials}
\end{equation}
we arrive at 
\begin{equation}
\begin{split}
    &\sum_{n=0}^{\infty} C_n \Big\{ \cos{(2n+1)\theta}-\cos{(2n+3)\theta}  -  \\
    & \left. - \frac{8 \lambda (n+1) }{w}  \sin{2(n+1)\theta} -\frac{4 \lambda (n+1)}{w}  \left[\cos{2n\theta} \right. \right. \\ 
    & \left.  -\cos{2(n+2)\theta}\right] \Big\} = 0.
\end{split}   
\label{eq: something wierd}
\end{equation}
For natural $m$ and $n$, we have 
\begin{equation}
\begin{aligned}
& \int_0^\pi \cos{(nx)} \cos{(mx)} dx = \frac{\pi}{2}\delta_{m,n}\ (\text{or } \pi \delta_{m,0} \; \text{for} \; n=0), \\
& \int_0^\pi \cos{(nx)} \sin{(mx)}dx = \frac{[(-1)^{n+m}-1]n}{m^2-n^2}.
\end{aligned}
\end{equation} 
We can multiple Eq.~\eqref{eq: something wierd} by $\cos{(m \theta)}$ and integrate it over $\theta$ from $0$ to $\pi$.

By using even $m$, we get
\begin{equation}
    \begin{aligned}
     \left(\frac{m}{2}+1\right)C_{\frac{m}{2}} & =\left(\frac{m}{2}+3\right)C_{\frac{m}{2}+2},\\
     C_0 & = 0.
    \end{aligned}
    \label{eq: even m}
\end{equation}
For odd values of $m$, we obtain
\begin{equation}
     \frac{\pi}{2}\left(C_{\frac{m-1}{2}}  - C_{\frac{m-1}{2}-1}\right)+\frac{32 \lambda}{ w}\sum_{n=0}^{\infty} C_n \frac{(n+1)^2}{4(n+1)^2-m^2} = 0.
    \label{eq: odd m}
\end{equation}
Since $m$ could be any natural number, we can safely establish the following relations for any integer $k$ and for any odd $m $:
\begin{equation}
\begin{split}
 C_{k+2} & = \frac{k+1}{k+3} C_k, \\
 C_{2k}=0, & \quad  C_{1+2k}=\frac{C_1}{k+1}, \\
 \left(\frac{\pm \pi}{m \pm 1}+\frac{128 \lambda}{ w}\right. & \left. \sum_{k=0}^{\infty} \frac{(k+1)}{16(k+1)^2-m^2} \right) C_1=0. 
\end{split}
\label{eq:coeffitients of expansion}
\end{equation}
Here the sign is $+$ if $(m-1)/2$ is even and $-$ if it is odd.
From the divergence of the sum in the last line of Eq.~\eqref{eq:coeffitients of expansion} we can conclude that the only meaningful solution has $C_1=0$ and, as follows from the second line of \eqref{eq:coeffitients of expansion}, $C_n=0$ for all $n$. Therefore $v_x(x, 0) = 0$ on the walls.

We should note here, that the Chebyshev polynomials expansion $f(x)=\sum_n^{\infty} C_n U_n(x)$ works only for functions with a finite integral $\int_{-1}^1 f(x) dx$. If we have a divergent, but integrable solution for $v_x$, which diverges as $1/\sqrt{1-\alpha^2}$ or stronger near the endpoints $\alpha = \pm 1$, then we cannot expand it in the way we did in Eq.~\eqref{eq:new form of v_x}. Strictly speaking, there could exist nontrivial integrable divergent solutions for $v_x(x,0)$ with the sign changes on the wall.

\clearpage

\setcounter{equation}{0}
\setcounter{figure}{0}
\global\long\def\theequation{S\arabic{equation}}
\global\long\def\thefigure{S\arabic{figure}}

\begin{widetext}
\begin{center}
\textbf{\large Supplemental Information for ``Viscous flow through a finite-width slit: Boundary conditions and dissipation''\\}
\vspace{15pt}
Daniil Asafov, Valentin Kachorovskii, Konstantin Tikhonov and Gu Zhang\\
\end{center}
\vspace{10pt}

Here we  provide a detailed derivation of Eq.~(C6) for a general  case of continuous velocity profile $v_y (x, y = 0)$  in our geometry. To begin with, we notice that  for any finite $w$, we have an identity
\begin{equation}
\begin{aligned}
& \int_{-\infty}^\infty dk_1 \frac{\sin[(k-k_1)w]}{k-k_1} e^{ik_1 x}  
=  \frac{\pi}{2} e^{ikx} [\text{sgn}(w-x) + \text{sgn}(w+x)].
\end{aligned}
\label{eq:identity_one}
\end{equation}
%As we have discussed above, both sides of the equation above are of order zero, once expressed in a dimensionless form.
We denote the Fourier transform of $\mathcal{A}_1 (k)$ as $\tilde{\mathcal{A}}_1 (x)$. 
Its derivative $\partial_x \tilde{\mathcal{A}}_1 (x)$ [following Eq.~(E3)] equals $v_y (x,0)$, i.e., the velocity along $y$ direction at $y=0$.
Then, the left-hand side of Eq.~(C6) can be expressed as
\begin{equation}
\begin{aligned}
& \int_{-\infty}^\infty \frac{dk_1}{ k - k_1 } \sin[(k-k_1) w]  k_1^2 \mathcal{A}_1 (k_1) 
=   \int_{-\infty}^\infty dx \tilde{\mathcal{A}}_1 (x) \left( - \frac{\partial^2}{\partial x^2} \right) \int_{-\infty}^\infty dk_1 \frac{\sin[(k-k_1)w]}{k-k_1} e^{-ik_1 x}  \\
& =\int_{-\infty}^\infty\! dx  \left\{  \frac{\pi}{2} k^2 [\text{sgn}(w + x) + \text{sgn}(w - x)] 
\right.\\
& \left. \ \ \ \ -2i\pi k [\delta (x-w ) - \delta (x + w)] 
  + \pi [ \delta'(x - w) - \delta'(x+w) ] \right\}e^{-ikx} \tilde{\mathcal{A}}_1(x).
\end{aligned}
\label{eq:k12_integral}
\end{equation}
For a continuous velocity profile, the velocity at the barrier end points equals zero, hence $\partial_x \tilde{\mathcal{A}}_1 (x) |_{x = \pm w}={v_y (\pm w,0)=0}$. In this case, the contribution to Eq.~\eqref{eq:k12_integral} of the terms with the derivatives of the delta functions becomes (through integration by parts):\\
$ i \pi k \left[ e^{-i k w}  \tilde{\mathcal{A}}_1 (w)- e^{ik w} \tilde{\mathcal{A}}_1 (-w) \right] $.

In addition, the expression for $\tilde{\mathcal{A}}(x)$ can be written as
\begin{equation}
\begin{aligned}
\tilde{\mathcal{A}}_1(x) = \int_{-\infty}^x dx' v_y(x',0) = \begin{cases}
0,\quad & \text{if } x<-w,\\
\int_{-w}^x dx' v_y (x',0),\quad & \text{if } -w<x<w,\\
I \frac{w}{e N},\quad & \text{if } x>w,
\end{cases}
\end{aligned}
\label{eq:c1x}
\end{equation}
where $I$ refers to the current through the slit. 
Note that $\tilde{\mathcal{A}}_1(-w)=0$
and $\tilde{\mathcal{A}}_1(w)=I w/(e N)$ independently of the profile of $v_y$ inside the slit.
The function $\tilde{\mathcal{A}}_1(x)$ for $-w<x<w$ does depend on this profile.
In particular, for our regular solution with $v_y(x,0)\propto \sqrt{1-x^2/w^2}$, the integral
 in Eq.~(\ref{eq:c1x}) reads:
\begin{equation}
    \int_{-w}^x dx' v_y (x',0) = I \frac{w}{e N} \left\{ \frac{1}{2} + \frac{1}{\pi} \left[ \frac{x}{w} \sqrt{1 - \frac{x^2}{w^2}} + \arcsin \left( \frac{x}{w} \right) \right] \right\}.
\end{equation}
Equation~\eqref{eq:c1x} leads to the expression of the Fourier transform
\begin{equation}
\begin{aligned}
\mathcal{A}_1(k) & = \int_{-\infty}^\infty dx e^{-ikx} \tilde{\mathcal{A}}_1(x)  
= \int_{-w}^w dx\  \tilde{\mathcal{A}}_1(x) e^{-ikx} +\frac{I w}{e N} \int_w^\infty e^{-ikx } dx
\end{aligned}
\label{eq:c1k_ft}
\end{equation}
and, hence,
\begin{equation}
\int_{-w}^w dx\   \tilde{\mathcal{A}}_1 (x) e^{-ikx}=\mathcal{A}_1(k)+i \frac{I w}{k e N} e^{-ikw}.
\end{equation}
With these facts, the integral of Eq.~\eqref{eq:k12_integral} then reduces to
\begin{equation}
\begin{aligned}
& \int_{- w}^w dx \frac{\pi}{2} k^2  e^{-ikx} \tilde{\mathcal{A}}_1 (x) - i \pi k \left[e^{-i k w}  \tilde{\mathcal{A}}_1 (w) - e^{ik w} \tilde{\mathcal{A}}_1 (-w) \right] \\
= &  \pi k^2 \int_{-w}^w dx e^{-ikx} \tilde{\mathcal{A}}_1(x) -  i\pi k e^{-ikw} \frac{I w}{e N}
=  \pi k^2 \mathcal{A}_1(k).
\end{aligned}
\label{eq:identity_three}
\end{equation}
We have thus arrived at Eq.~(C6).

Derivations above do not rely on the explicit expression of $v_y (x,0)$ in the slit.
For our specific situation, we can derive the expression of $\mathcal{A}_1 (k)$ using the explicit expression of $v_y (x,0)$ [obtained from Eq.~(12) of the main text]. 
The analytic expression of $\mathcal{A}_1 (k)$ reads:
\begin{equation}
    \mathcal{A}_1 (k) =-i \frac{2I}{e N} \frac{J_1( k w )}{k^2}.
    \label{eq:c1k_explicit}
\end{equation}
Following Eq.~\eqref{eq:c1k_explicit}, we notice that $\mathcal{A}_1 (k)$ is $w$-dependent.

\end{widetext}

\end{document}